\pdfoutput=1

\documentclass[%
 prr,
 amsmath,amssymb,
 reprint,%
 twocolumn,
 superscriptaddress
]{revtex4-2}
\usepackage[normalem]{ulem}
\usepackage{filecontents}
\usepackage[version=3]{mhchem}
\usepackage{color} 
\usepackage{siunitx}

\usepackage{multirow}
\usepackage{makecell}
\usepackage{subcaption}
\usepackage{relsize}

\usepackage{tikz}
\usepackage{mathtools}
\usetikzlibrary{arrows.meta, decorations.pathreplacing,decorations.markings, decorations.pathmorphing, positioning, shapes, calc,fadings}
\tikzfading 
[
name=fade out,
inner color=transparent!0,
outer color=transparent!100
]
\tikzset{snake it/.style={decorate, decoration=snake}}
\usepackage{tikz-feynhand}

\usepackage{graphicx}
\usepackage{dcolumn}
\usepackage{bm}

\usepackage[utf8]{inputenc}
\usepackage[T1]{fontenc}

\usepackage{mathptmx}
\DeclareSymbolFont{newfont}{OML}{cmm}{m}{it}
\DeclareMathSymbol{\epsilon}{3}{newfont}{15}

\usepackage{xr}
\makeatletter
\newcommand*{\addFileDependency}[1]{
  \typeout{(#1)}
  \@addtofilelist{#1}
  \IfFileExists{#1}{}{\typeout{No file #1.}}
}
\makeatother

\newcommand*{\myexternaldocument}[1]{%
    \externaldocument{#1}%
    \addFileDependency{#1.tex}%
    \addFileDependency{#1.aux}%
}
\myexternaldocument{./__bsupp}
\usepackage{prettyref}
\newrefformat{supp-fig}{S\ref{#1}}
\newrefformat{supp-tbl}{S\ref{#1}}
\newrefformat{supp-eq}{S\ref{#1}}
\DeclareUnicodeCharacter{2212}{-}

\newcommand{\mbf}[1]{\mathbf{#1}}

\usepackage{lipsum}

\begin{document}

\preprint{AIP/123-QED}
\title[]{\textcolor{black}{Anti-Coulomb ion-ion interactions: a theoretical and computational study}}

\newcommand{\SBUphysast}{Physics and Astronomy Department, Stony Brook University, Stony Brook, New York 11794-3800, United States}
\newcommand{\IACS}{Institute for Advanced Computational Science, Stony Brook University, Stony Brook, New York 11794-3800, United States}
\newcommand{\UADM}{Departamento e Instituto de Física de la Materia Condensada (IFIMAC), Universidad Aut\'{o}noma de Madrid, E-28049 Madrid, Spain}

\author{Alec Wills}
\email{alec.wills@stonybrook.edu}
\affiliation{\SBUphysast}
\affiliation{\IACS}
\author{Anthony Mannino}
\affiliation{\SBUphysast}
\affiliation{\IACS}
\author{Isidro Losada}
\affiliation{\UADM}
\author{Sara G. Mayo}
\affiliation{\UADM}
\author{Jose M. Soler}
\affiliation{\UADM}
\author{Marivi Fern\'{a}ndez-Serra}%
 \email{maria.fernandez-serra@stonybrook.edu}
\affiliation{\SBUphysast}
\affiliation{\IACS}

\date{\today}

\begin{abstract}

The free energy of ion solvation can be decomposed  into enthalpic and entropic contributions. This helps to understand the connection between the dielectric properties and the underlying forces. We present a simple linear-response model of screened charge interactions that provides an alternative understanding of solvation barriers. Moreover, it explains the ``anti-Coulomb'' interactions (attraction between like-charged ions and repulsion between opposite-charged ions) observed in both simulations and experiments. We show that this is a universal behavior associated to the non-local response function of any dielectric or metallic system.

\end{abstract}

\maketitle

\section{\label{sec:level1} Introduction}

The study of electrolyte solutions is of paramount importance to a wide variety of natural and industrial processes, including human biology, advancements in material design, and interfaces between solution and solid \cite{rev1Anderko2002,rev2Shcherbakov2021,molinero_DeMille2011_mwiondna,molinero_Lu2017_aem,molinero_Lu2019_aem2, selloni1_Ding2020}.
In principle, solvation is controlled by simple, classical electrostatic interactions \cite{model3Kharkats1976,model1Wolynes1980,model2Friedman1981}.
In practice, however, simulations with small changes in the interatomic forces and modeling methods return quantitatively and qualitatively different results \cite{Wills2021, rev3PascualP2022,edl1Baldelli2008,edl2Wang2016, molinero_DeMille2009_mwion,clsolvdft_DelloStritto2020}, and screening effects can lead to very counterintuitive behaviors in electrolyte solutions.
Thus a thorough analysis of solvent-solute interactions is required to understand such solutions.
Computational and theoretical studies that can separate the different interactions and thermodynamic quantities will clarify these processes.

Of special interest are ion-ion and ion-solvent interactions in water, and solvated sodium chloride has been a prototypical subject of study.
Some studies have shown counterintuitive attractive states of same-charge ion-ion pairs.
The existence of a contact ion pair (CIP) has been predicted for \ce{Cl- - Cl-} \cite{Dang1987_oldii1,Dang1990_oldii2,Dang1992_oldii3}, although with varying degrees of stability or even with no stable state \cite{Guardia1991_iidyn}.
For \ce{Na+ - Na+} pairs, some local minima were found \cite{Guardia1991_iidyn} but without a stable CIP state \cite{Dang1987_oldii1,Dang1990_oldii2, Dang1992_oldii3}.
In the electric double layer of water between two electrodes, attraction was found between divalent counterions, but not between monovalent ones \cite{Guardia1991_iidyn}.
Further simulations with better forces and computational power also showed stable states of \ce{Cl- - Cl-} and \ce{Na+ - Na+} pairs in water \cite{Keasler2006_watermediatedii}.
Reference site models have been used to characterize the interactions and the bound states of these pairs \cite{Kovalenko2000_rism1,Kovalenko2000_rism2}.
Furthermore, colloid suspension experiments using bright-field optical microscopy have found that like-charged particles can attract and oppositely-charged particles can repel, and that this can be used to drive cluster self-assembly \cite{beadpairing_Wang2024}.
\textcolor{black}{
Theoretical studies investigating the interaction energies of spheres and planes in dielectric media have shown that induced polarization of bound charges can lead to attraction between like-charges \cite{screenref12_Khachatourian2014}.
However for two spheres in a dielectric, this attraction was only seen at specific ratios of the sphere charges and radii \cite{screenref11_Bichoutskaia2010,screenref13_Lindgren2016,screenref15_Chan2020}, implying asymmetry is required for the attraction to exist.
The polarization characteristics of the solute and solvent have shown to be important factors in driving self-assembly of nano-materials through these like-charge attractions \cite{screenref14_Lindgren2018}.
}

Molecular configurations for these stable states have been proposed, often treating the water molecules as a stabilizing bridge between two anions \cite{Zangi2012_iimono}.
Neural network and experimental diffraction studies have found extended effects on water shells around solute ions \cite{Soper2007_extended,Zhang2022}, and experimental support for the anionic bridging was also found \cite{Soper2007_chaochosmo}.
There are also continuous-solvent explanations of this counterintuitive binding.
In such formalisms, an inversion of the sign of the dielectric function ($\epsilon < 0$) leads to \textit{overscreening}, or effective repulsion between unlike charges and attraction between like charges \cite{Kornyshev1997_overscreening}.
This is the same mechanism behind phonon-mediated 
attractive electron-electron interaction in metals \cite{Ashcroft76,Allen1988_totaldiel}.

\begin{figure*}[ht!]
     \centering
     \begin{subfigure}[b]{0.32\textwidth}
     \includegraphics[scale=0.34]{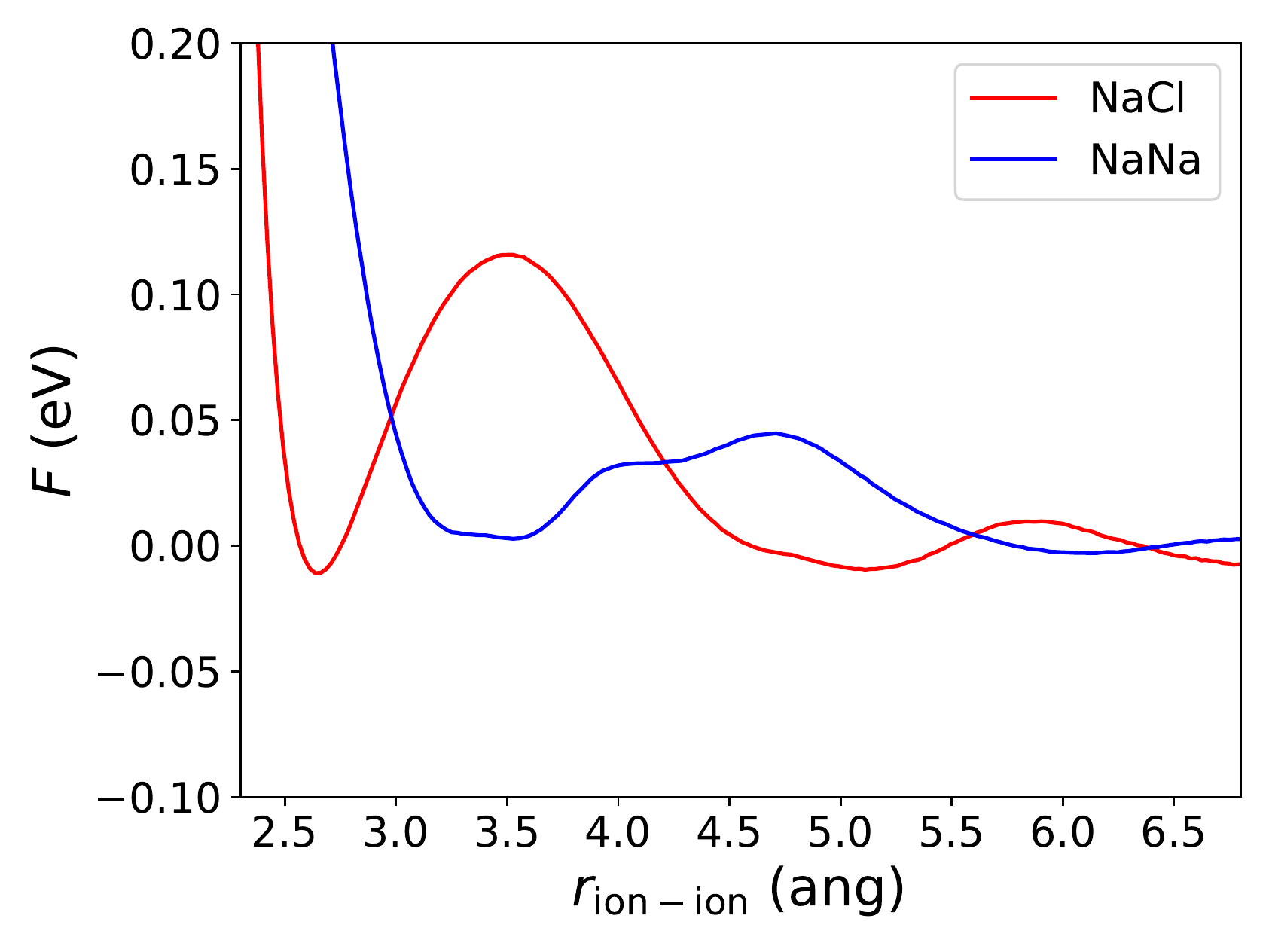}
         \caption{}
         \label{fig:spceopls-pmf}
     \end{subfigure}
     \begin{subfigure}[b]{0.32\textwidth}
     \includegraphics[scale=0.34]{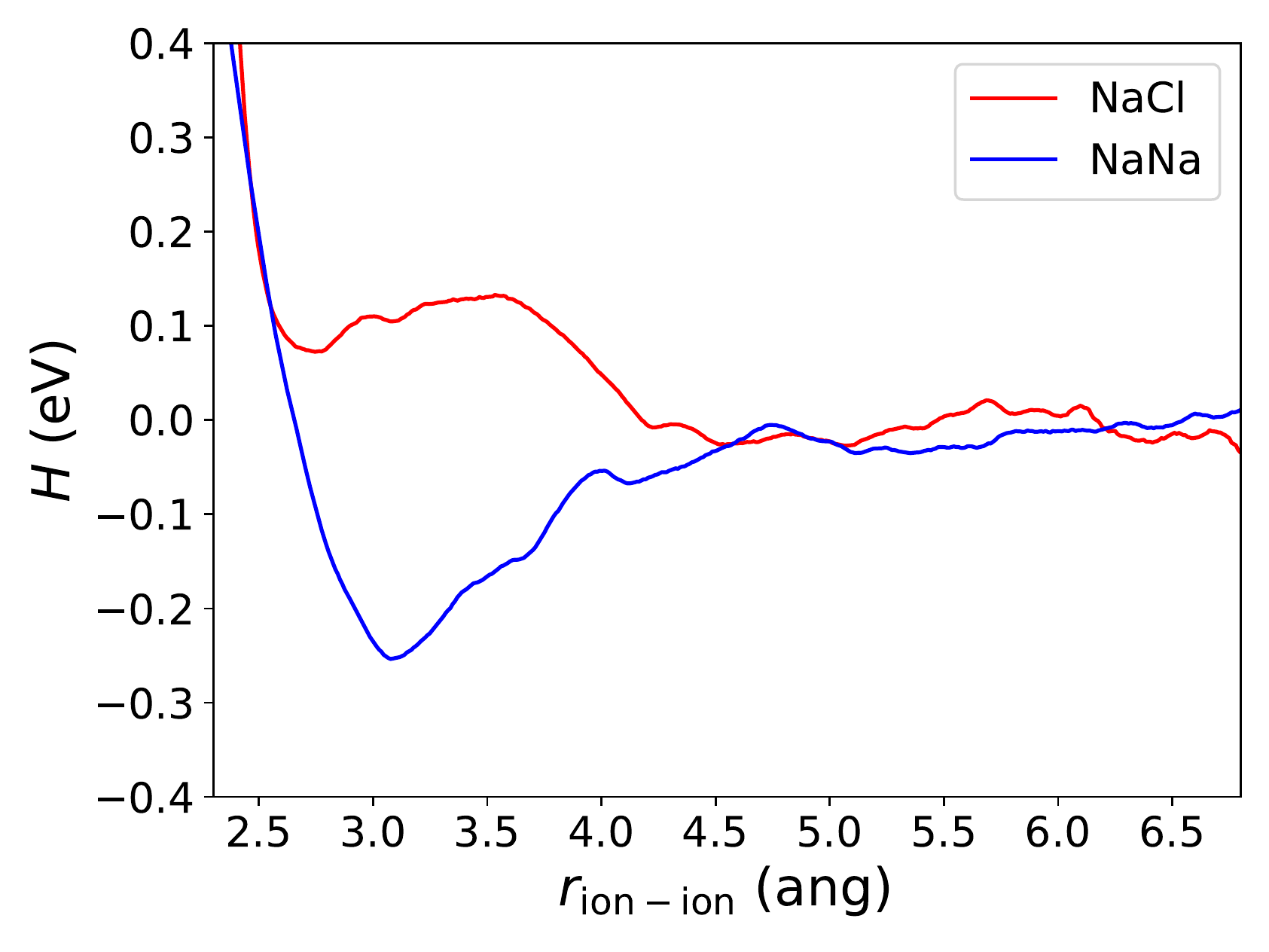}
         \caption{}
         \label{fig:spceopls-pot}
     \end{subfigure}
     ~ 
     \begin{subfigure}[b]{0.32\textwidth}
     \includegraphics[scale=0.34]{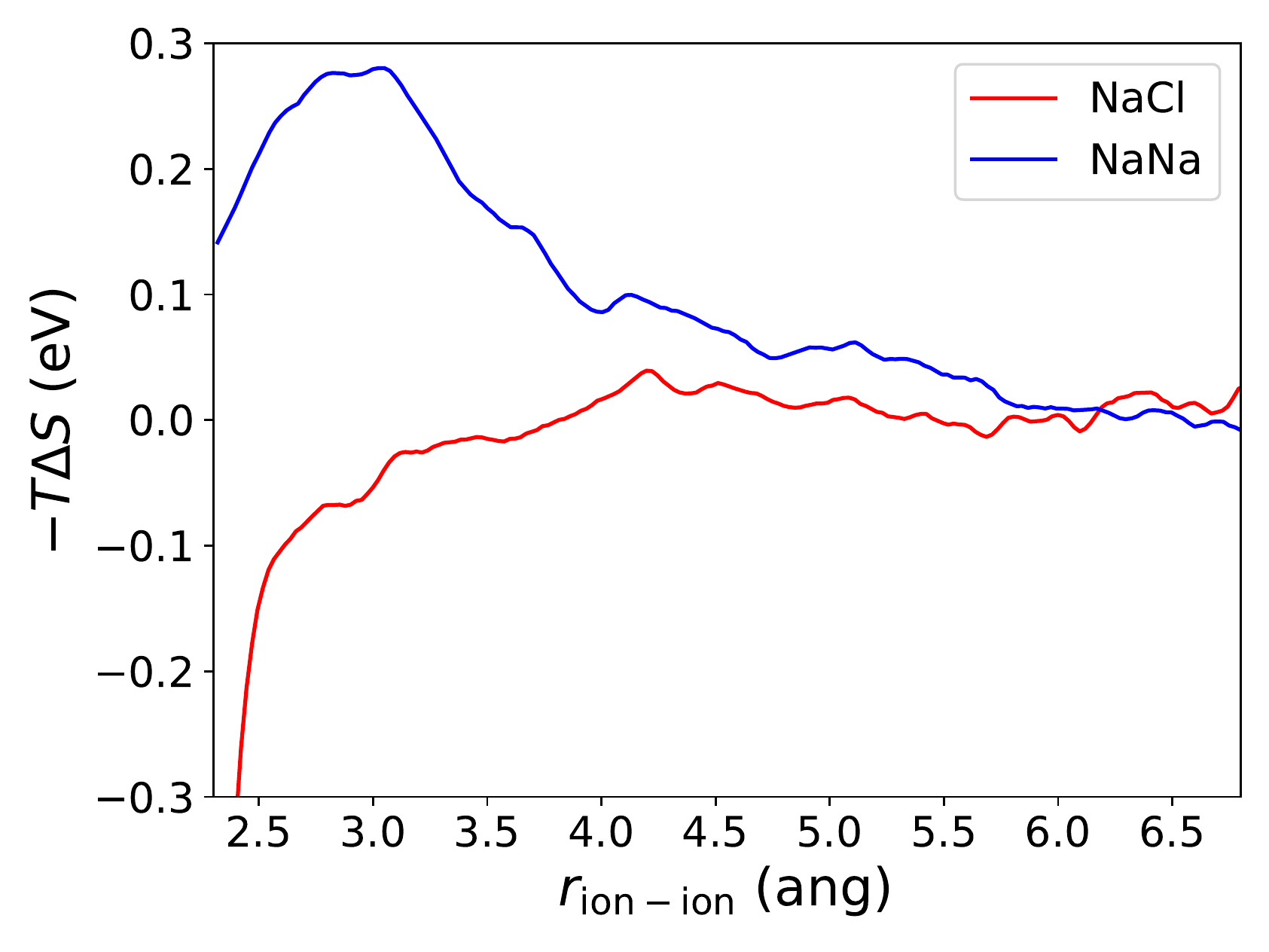}
         \caption{}
         \label{fig:spceopls-tds}
     \end{subfigure}
     \caption{\textbf{(a)} Potentials of mean force for \ce{Na+ - Cl-} (red) and \ce{Na+ - Na+} (blue) ions, solvated in water (with the SPC/E model), as function of interionic distance. \textbf{(b)} Mean potential energies, giving the enthalpic contribution to the free energy. \textbf{(c)} Entropic contribution to the free energy, taken as $-T\Delta S = F - H$. The zeros for $F$ and $H$ were taken as their average for $r>5.5$ \AA.}
\end{figure*}

Further, decomposition of the free energy into enthalpic and entropic contributions shows counterintuitive enthalpy barriers between oppositely-charged ions \cite{galli_Zhang2020,Wills2021}.
This underscores the importance of understanding solute-solvent interactions and the free energy landscape.
In the first part of this work, we confirm these previously mentioned, unexpected occurrences in simulations of \ce{Na+ - Na+} and \ce{Na+ - Cl-} pairs in water.
Thus, we find that transition barriers between opposite-charge ions are robust across different simulated systems.
%
Moreover, the presence of stable states between same-charge ions is also validated.

In the second part of this work, we introduce a very simple but highly counter-intuitive model of electrostatic interactions in dielectric media.
This model provides a straightforward explanation of stable states of 
\ce{Na+ - Na+} and \ce{Cl- - Cl-} pairs, and of a repulsive barrier of \ce{Na+ - Cl-} pairs in water.
These predictions do not need the sign inversion of the dielectric constant that leads to overscreening, \textcolor{black}{nor does it require asymmetry between the particles for attraction to exist.}
Thus, combining theory and simulations, we offer a new perspective of the physics and processes of salt solvation.

\section{Charges in Water: \ce{Na+ - Na+} vs. \ce{Na+ - Cl-}}

In order to study the interaction between like-charge and opposite-charge ions, we compute the potential of mean force (PMF) $U$, as a function of the interionic separation $r$.
It can be succinctly stated \cite{pmfdefKirkwood1935}  as $$U(r) = -kT\ln g(r), \label{eq:pmfdef}$$ where $k$ is Boltzmann's constant, $T$ is the temperature, and $g(r)$ is the ion-ion radial distribution function.
This in principle gives us a straightforward way to calculate $U(r)$ from unconstrained simulations.
But the solvation landscape strongly favors specific states over others, making a converged $g(r)$ hard to calculate efficiently.
Also, hydration shells can be very long-lasting and strongly defer convergence \cite{Li2008}.
For these reasons, constrained simulations are necessary in practice to force the exploration of configuration space.
However, the introduction of constraints requires additional work, such as choices and tests of sampling methods and reaction coordinate discretization, among many others, to ensure converged, repeatable results.
Recent work has reviewed these choices and their effects in detail, to which we direct the reader \cite{Wills2021}.

\subsection{Methodology}

Due to the ease of decomposing the total potential energy into constituent pairwise contributions, our classical molecular dynamics (MD) simulations are performed with the \textsf{GROMACS} \cite{gromacsAbraham2015} software suite in the NVT ensemble.
A cubic box with a side length of 14.373 \AA\ was filled with 96 water molecules and two ions, either \ce{Na+ - Cl-} or \ce{Na+ - Na+}.
To sample the PMF, harmonic restraints were placed at reference interionic distances ranging from 2.0 to 6.8 \AA\ in steps of 0.1 \AA. 
Eight different random seeds were used to generate separate dynamics during equilibration, after which production runs with $\Delta t=\SI{0.5}{fs}$ were run for $2\times10^6$ steps, yielding 1 ns of sampling data per random seed and restraint distance (8 ns per reference distance).
Additionally, for the same eight random seeds, a lone Na$^+$ or Cl$^-$ ion was simulated in the same box of 96 water molecules to use as the reference in the infinitely dilute limit.

For these classical simulations, we used a variety of water models and ion parameters.
The figures in the main body of this paper were generated with the SPC/E \cite{spce_Berendsen1987} water model and OPLS-AA ion parameters \cite{oplsJorgensen1996}.
In the supplementary section, we show results from various combinations of different water models, such as TIP4P and TIP4P/2005, and a modified pair of ion parameters for interionic constraints from 2.0 \AA\ to 6.0 \AA\ \cite{tip4pJorgensen1983, t4p05Abascal2005,naclt4pew_Joung2008}, all showing the expected behavior to varying degrees.
\textcolor{black}{
Furthermore, to investigate the effects that system size and polarizability play in these interactions, we conducted simulations in \textsf{LAMMPS} \cite{LAMMPSnew} using varying combinations of polarizable models for the ions and water \cite{swm4ndpLamoureux2006,swm4_ions_Yu2010}, with systems of varying sizes and number of ions.
}

\subsection{Free Energy Landscape Decomposition}

\begin{figure}[t!]
    \centering
    \includegraphics[scale=0.5]{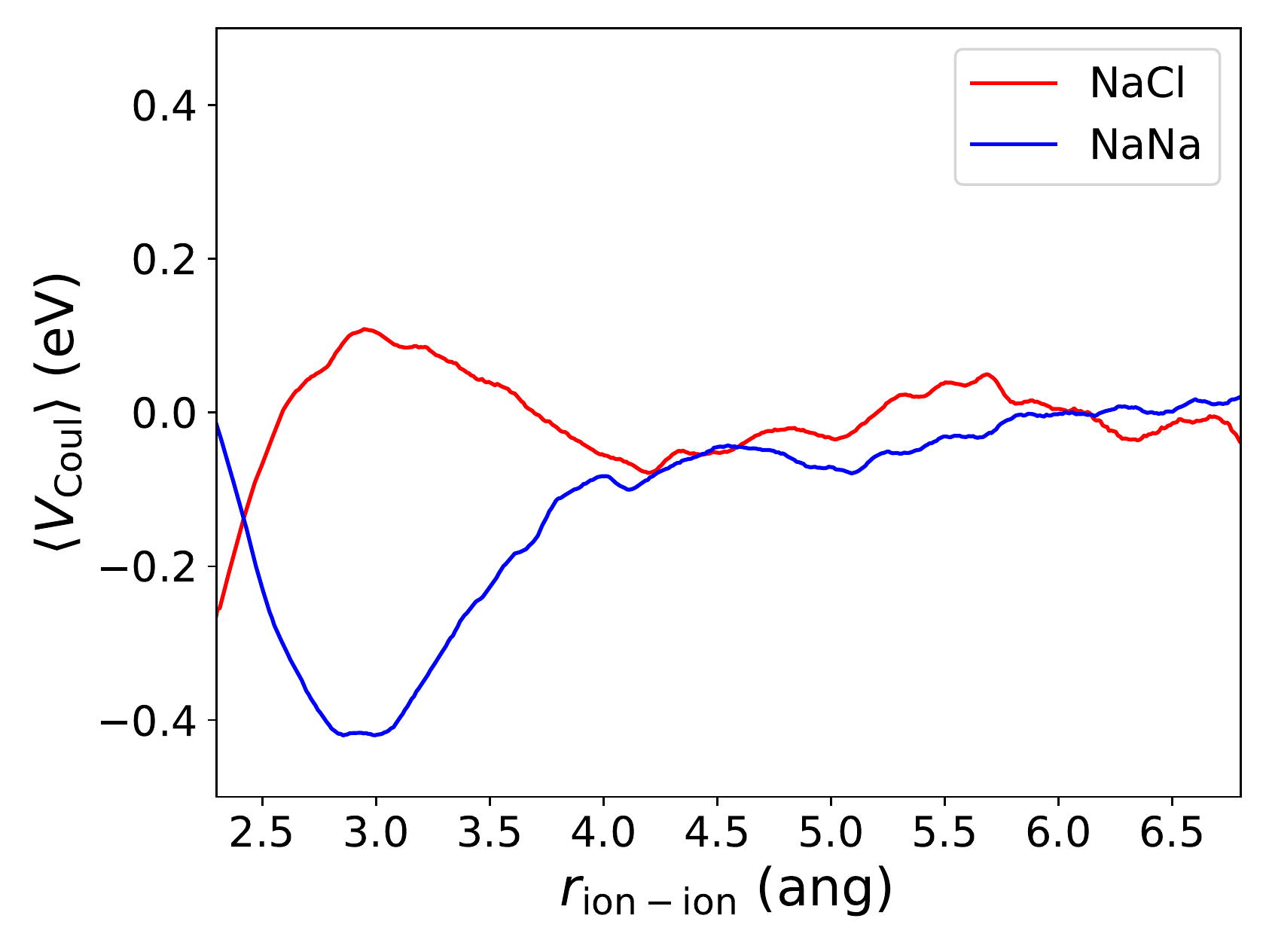}
    \caption{Average Coulomb energy as a function of interionic distance. A barrier between oppositely charged ions (\ce{Na+ - Cl-}, red) is clearly visible as the ions are brought together. On the other hand, there is a strong attraction between the two like-charged ions (\ce{Na+ - Na+}, blue). The zero value was taken as the average beyond 5.5 \AA.}\label{fig:spceopls-coul}
\end{figure}

The PMF $F$ was generated using the weighted histogram analysis method (WHAM) \cite{whamKumar1992}, as implemented in the \textsf{wham} program \cite{whamGrossfield}.
It is shown in Fig.~\ref{fig:spceopls-pmf} as a function of interionic separation.
The average enthalpy $H$ is shown in Fig.~\ref{fig:spceopls-pot}, and a similar average of just the Coulomb energy of the system is shown in Fig.~\ref{fig:spceopls-coul}.
In both figures, we see clear validation of previous results: energetically bound states between like-charge cations and repulsive barriers between opposite-charge ions.
The entropic contribution to the free energy $-T\Delta S=F-H$ is shown in Fig~\ref{fig:spceopls-tds}.
Notice that, since $T>0$, the changes in $\Delta S$ are opposite to those shown in the figure.
The bound state of the cation pair is clearly stabilized by the electrostatic enthalpy, despite the entropy decrease as the pair is brought together.
On the other hand, for the \ce{Na+ - Cl-} pair, the contact minimum at $r\simeq 2.5 \AA$ is stabilized by the increase in entropy.

It is interesting to note that not only the electrostatic interaction but also the entropy changes sign when comparing same- and opposite-charge pairs.
While Fig.~\ref{fig:spceopls-tds} shows that, as two sodium cations are brought together, the entropy decreases until $r\simeq 3.0 \AA$, in the case of the \ce{Na+ - Cl-} pair it monotonically increases
with shorter inter-ionic distances.
For \ce{Na+ - Na+} this indicates that interstitial water molecules forming the solvation shell have larger entropy than when they are in the liquid bulk, while the opposite applies to \ce{Na+ - Cl-}.

The simulation's electrostatic energy is shown in Fig.~\ref{fig:spceopls-coul}.
As the like-charge and opposite-charge ion pairs are brought closer together, the screening effects modify the effective electrostatic potential to yield an energetically favored region of attraction and an unfavorable repulsive region, respectively.
Indeed, we see that these features persist in the enthalpy decomposition shown in Fig.~\ref{fig:spceopls-pot}.
Interestingly, for the water model and ion parameter combination shown, the enthalpic stability for the \ce{Na+ - Na+} pair allows for a weakly stable state in the free energy profile.

\section{Charges in a Dielectric: A Simple Model}

The calculation of the interaction energy between two charges embedded in a dielectric involves computing the medium's response to these charges, typically through its non-local dielectric function $\epsilon(\mbf{r},\mbf{r}')$.
We compute this interaction from the linear response to a generic external charge density, from which one can calculate the perturbing potential.

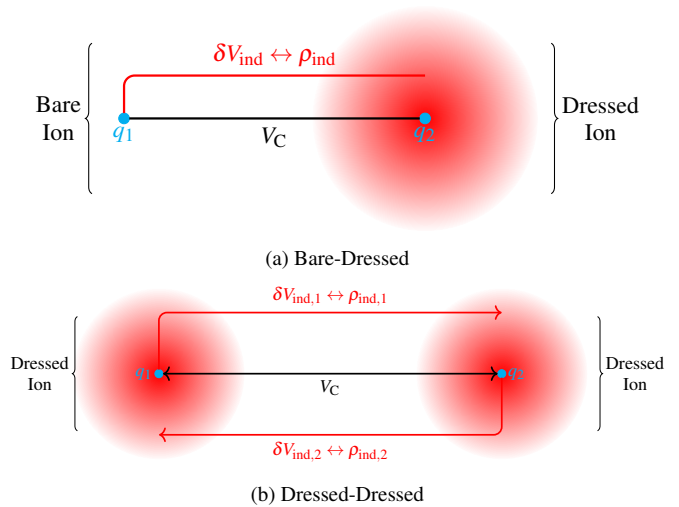
\begin{figure}[t!]
    \begin{subfigure}[b]{0.5\textwidth}
    \centering
    		\begin{tikzpicture}
		
		\filldraw[cyan] (0,0) coordinate (q1) circle(2pt) node[below]{$q_1$};
		
		\tikzset
		{
			myCircle/.style=
			{
				red,
				path fading=fade out,
			}
		}
		\fill[myCircle] (4,0) circle (1.5);
		\filldraw[cyan] (4,0) coordinate (q2) circle(2pt) node[below]{$q_2$};
		(q2)++(-2pt,0) coordinate (q2m)
		\draw[decoration={brace,mirror,raise=5pt},decorate]
		(-0.25,1) -- node[align=center, left=6pt] {Bare\\Ion} (-0.25,-1);
		\draw[decoration={brace,mirror,raise=5pt},decorate]
		(5.5,-1) -- node[align=center, right=6pt] {Dressed\\Ion} (5.5,1);
		\draw[line width=0.3mm] (q1)++(2pt,0) -- ($ (q2) + (-2pt,0) $) node[midway,below]{$V_\mathrm{C}$};
		\draw[rounded corners, line width=0.3mm, red] (q1)++(0,2pt) -- ++(0,0.5) -- ($ (4,0.5) + (0,2pt) $) node[midway,above]{$\delta V_\mathrm{ind}\leftrightarrow \rho_\mathrm{ind}$};
		\end{tikzpicture}
		\caption{Bare-Dressed}
		    \label{fig:baredressed}
        \end{subfigure}
        \begin{subfigure}[b]{0.5\textwidth}
            \centering
            \resizebox{\textwidth}{!}{
			\begin{tikzpicture}
			\tikzset
			{
				myCircle/.style=
				{
					red,
					path fading=fade out,
				}
			}
			
			\fill[myCircle] (q1) circle (1.5);
			\filldraw[cyan] (0,0) coordinate (q1) circle(2pt) node[left]{$q_1$};
			\fill[myCircle] (6,0) circle (1.5);
			\filldraw[cyan] (6,0) coordinate (q2) circle(2pt) node[right]{$q_2$};
			(q2)++(-2pt,0) coordinate (q2m)
			\draw[line width=0.3mm, <->] (q1)++(2pt,0) -- ($ (q2) + (-2pt,0) $) node[midway,below]{$V_\mathrm{C}$};
			\draw[rounded corners, line width=0.3mm, red, ->] (q1)++(0,2pt) -- ++(0,1) -- ($ (6,1) + (0,2pt) $) node[midway,above]{$\delta V_\mathrm{ind,1}\leftrightarrow \rho_\mathrm{ind,1}$};
			\draw[rounded corners, line width=0.3mm, red, ->] (q2)++(0,-2pt) -- ++(0,-1) -- ($ (0,-1) + (0,-2pt) $) node[midway,below]{$\delta V_\mathrm{ind,2}\leftrightarrow \rho_\mathrm{ind,2}$};
			\draw[decoration={brace,mirror,raise=5pt},decorate]
			(-1.25,1) -- node[align=center, left=6pt] {Dressed\\Ion} (-1.25,-1);
			\draw[decoration={brace,mirror,raise=5pt},decorate]
			(7.5,-1) -- node[align=center, right=6pt] {Dressed\\Ion} (7.5,1);
			\end{tikzpicture}
			}
			\caption{Dressed-Dressed}
			\label{fig:dresseddressed}
        \end{subfigure}
    \caption{\textbf{(a)} A bare ion test charge $q_1$ to measure $\delta V_\mathrm{ind}$ induced by $q_2$ in a dielectric medium, where we take the external perturbing potential to be the Coulomb potential of $q_2$. Its induced potential is schematically shown as the cloud around $q_2$. We call the perturbing ion and its induced potential a ``dressed ion." Connecting lines indicate additive potentials. \textbf{(b)} Two bare ions with their first-order induced potentials. In the dressed-dressed interaction, the Coulomb potential is supplemented by each induced potential surrounding the constituent ion.}
    \label{fig:bddd}
\end{figure}

\subsection{Linear Response}

For completeness, we derive here the standard linear response equations \cite{Ashcroft76} that will be used to compute the interaction between charged particles embedded in a dielectric medium.
The Poisson equation for an externally perturbing potential is $$-\nabla^2\phi_\mathrm{ext}(\mbf{r}) = 4\pi \rho_\mathrm{ext}(\mbf{r}),$$ where $\rho_\mathrm{ext}$ is the density of the perturbing particle, placed at the origin.
Another Poisson equation is satisfied by the \textit{total} potential $\phi$ and density $\rho$: $$-\nabla^2\phi(\mbf{r}) = 4\pi \rho(\mbf{r}),$$ where $\rho=\rho_\mathrm{ext}+\rho_\mathrm{ind}$ is the total charge density of both the perturbing particle and the induced screening density.

One assumes a linear medium, such that the external potential and total potential are linearly related through $$\phi_\mathrm{ext}(\mbf{r}) = \int d\mbf{r}' \epsilon(\mbf{r},\mbf{r}')\phi(\mbf{r}'),$$ where spatial homogeneity implies that $\epsilon(\mbf{r},\mbf{r}') = \epsilon(\mbf{r}-\mbf{r}')$. This further implies diagonality in reciprocal space: \begin{equation}\phi_\mathrm{ext}(\mbf{q}) = \epsilon(\mbf{q})\phi(\mbf{q}) \leftrightarrow \phi(\mbf{q}) = \frac{1}{\epsilon(\mbf{q})}\phi_\mathrm{ext}(\mbf{q}).\label{eq:phiextephi}\end{equation}

However, it can be more natural to work directly with the charge density induced in the dielectric medium ($\rho_\mathrm{ind}$) by the external potential. If $\rho_\mathrm{ind}$ and $\phi$ are also linearly related (as should be the case for a weak enough $\phi$), their Fourier transforms are also linearly related through $\chi$: \begin{equation}\rho_\mathrm{ind}(\mbf{q}) = \chi(\mbf{q})\phi(\mbf{q}),\label{eq:rhochiq}\end{equation} where $\chi$ is the susceptibility of the material.

To relate $\epsilon$ to $\chi$, we Fourier transform the Poisson equations above and, letting $q=|\mbf{q}|$, we find $$q^2\phi_\mathrm{ext}(q) = 4\pi \rho_\mathrm{ext}(q), \ \ q^2\phi(q)=4\pi\rho(q),$$ which, together with the linear response relations, give $$\frac{q^2}{4\pi} \left[\phi(q)-\phi_\mathrm{ext}(q) \right] = \chi(q)\phi(q) \leftrightarrow \phi(q) = \frac{\phi_\mathrm{ext}(q)}{1-\frac{4\pi}{q^2}\chi(q)},$$ yielding \begin{equation}
    \epsilon(q) = 1-\frac{4\pi}{q^2}\chi(q).
    \label{eq:etochi}
\end{equation}
Equipped with Eqs.~\eqref{eq:phiextephi}, ~\eqref{eq:rhochiq}, and ~\eqref{eq:etochi}, one just needs to specify an exact or approximate dielectric function $\epsilon$ to compute the effective interaction between charges in the dielectric. 

\begin{figure*}[!ht]
     \centering
     \includegraphics[scale=0.5]{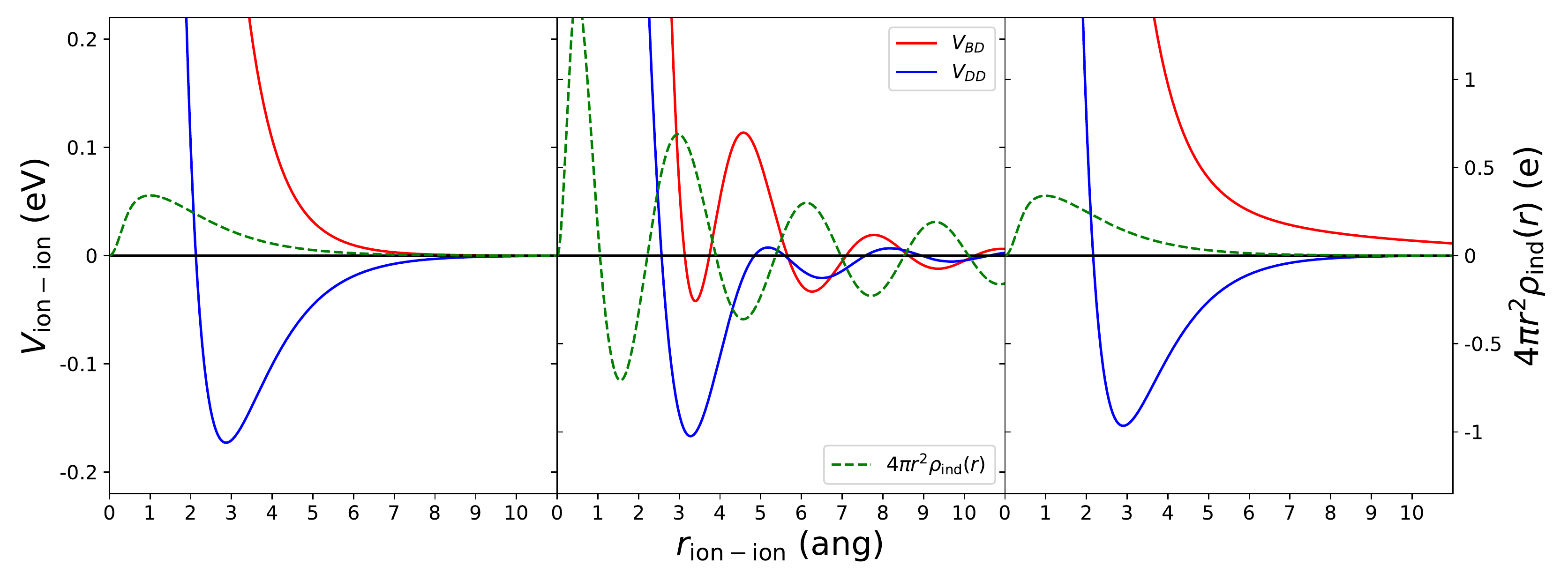}
     \caption{Bare-dressed (red) and dressed-dressed (blue) interaction energies (left y-axis) and induced densities (green, right y-axis) for the Thomas-Fermi approximation \textbf{(left)}, RPA approximation \textbf{(middle)}, and Inkson's dielectric function with $\epsilon_{k\to0}=80$ \textbf{(right)}. All plots are generated with $k_F=k_0=1$, using external Gaussian charge densities of width $\sigma_\mathrm{ext}=0.25$ \AA.}
     \label{fig:bdddfigs}
\end{figure*}

\subsection{Dressed Potentials}

The net effect of the screening, whether quantum mechanical (exchange and correlation) or electrostatic in origin, is to induce a surrounding charge around the perturbation source.
When the perturbing source is a point charge, it is natural to define a ``dressed particle" as a quasiparticle charged with the sum of the original and screening charge densities. 
A diagram depicting this scenario is given in Fig.~\ref{fig:bddd}, showing the various densities and potentials at play when considering interactions between pairs of these dressed or bare particles.

In the interacting electron gas, this dressing is nothing else but $n[g(r)-1]$, where $g(r)$ is the electron-electron pair-correlation function and $n$ is the average electronic density, $n=3e/(4\pi r_s^3)$ with $r_s$ being the Wigner-Seitz radius.
Overhauser originally showed that this distribution can be computed by solving the two electron scattering problem with an effective screened Coulomb repulsion, with later studies extending his analysis \cite{Overhauser1995,overhauser2Perdew2001,overhauser3}.
In this model, an electron scatters with the screened Coulomb potential generated by the other electron.
The effective interaction used by Overhauser is what we refer to here as the ``bare-dressed" interaction.
Later, Corona {\it et al.} proposed an alternative ``dressed-dressed" effective interaction to solve the same problem \cite{Corona2004}.
In such a description, the scattering occurs between two neutral particles -- each electron dressed by their corresponding exchange and correlation hole.

Here we argue that the interaction between two charges embedded in a dielectric medium corresponds to the interaction between two of such ``dressed quasiparticles," i.e. the Coulomb interactions between two charges and their respective dressings.
We describe here how we define the problem and obtain the inter-particle energies as a function of their separation.
In addition, this formulation of the problem can be readily adapted to the evaluation of inter-charge interactions in water (or other dielectric liquids), given that the pair correlation functions are directly obtained from molecular dynamics simulations.

\paragraph{Formalism:} With charge density sources $S_1(\mbf{r}_1)$ and $S_2(\mbf{r}_2)$ interacting through the Coulomb kernel $$K(\mbf{r},\mbf{r}') = \frac{\alpha_C}{|\mbf{r}-\mbf{r}'|},$$ where $\alpha_C$ is the proportionality constant for the Coulomb interaction, the interactions can be described by the integral \begin{equation}E[\mbf{r}_1,\mbf{r}_2] = \int d^3r d^3r' S_1(\mbf{r}-\mbf{r}_1)K(\mbf{r},\mbf{r}')S_2(\mbf{r}'-\mbf{r}_2)\label{eq:sks}\end{equation}
For an isotropic medium, $E[\mbf{r}_1,\mbf{r}_2]=E[\mbf{r}_1-\mbf{r}_2]=E[\mbf{r}]$.
These convolutions can be easily calculated in reciprocal space and Fourier-transformed back into real space.
Indeed, without access to a closed form of $\rho_\mathrm{ind}(r)$, it is necessary to use the $k$-space linear response relations in order to calculate the interactions involving the induced screening charge.

\paragraph{Bare-Bare Interaction}
For two point-charge sources $$S_i = Q_i\delta(\mbf{r}-\mbf{r}_i),$$ the integral in Eq.~\eqref{eq:sks} reduces to the regular Coulomb interaction: \begin{equation}E_\mathrm{BB}[\mbf{r}_1,\mbf{r}_2] = \alpha_C\cdot \frac{Q_1Q_2}{|\mbf{r}_1-\mbf{r}_2|}.\label{eq:ebb}\end{equation}

\paragraph{Bare-Dressed Interaction}
Here, one source is still an unscreened point charge (which we can think of as a test charge with which to measure the potential of the screened charge). The dressed charge will have a source term of the form $$S_i(\mbf{r}-\mbf{r}_i) = Q_i\delta(\mbf{r}-\mbf{r}_i) + \rho_\mathrm{ind, i}(\mbf{r}-\mbf{r}_i),$$ where $\rho_\mathrm{ind, i}$ is the induced screening charge from the medium. With these forms of source terms, we find in this case that \begin{align*}
    E_\mathrm{int}[\mbf{r}_1,\mbf{r}_2] &= E_\mathrm{BB}[\mbf{r}_1,\mbf{r}_2]\\
    &+ E_\mathrm{B_2 D_1}[\mbf{r}_1,\mbf{r}_2],
\end{align*} where \begin{equation}E_\mathrm{B_iD_j}[\mbf{r}_1,\mbf{r}_2] = \alpha_C Q_i\int d^3r \frac{\rho_\mathrm{ind,j}(\mbf{r}-\mbf{r}_j)}{|\mbf{r}-\mbf{r}_i|}\label{eq:ebd}\end{equation} is the interaction energy between the undressed point charge and the screening charge induced around the other point.

\paragraph{Dressed-Dressed Interaction}
Now, with each source having its own dressing, one finds upon expanding the terms in Eq.~\eqref{eq:sks} that \begin{align*}E_\mathrm{int}[\mbf{r}_1,\mbf{r}_2] &= E_\mathrm{BB}[\mbf{r}_1,\mbf{r}_2] \\&+ E_\mathrm{B_1D_2}[\mbf{r}_1,\mbf{r}_2]\\ &+ E_\mathrm{B_2D_1}[\mbf{r}_1,\mbf{r}_2]\\ &+ E_\mathrm{D_1D_2}[\mbf{r}_1,\mbf{r}_2], \end{align*} where \begin{equation}E_\mathrm{D_1D_2}[\mbf{r}_1,\mbf{r}_2] = \alpha_C \int d^3r d^3r' \frac{\rho_\mathrm{ind,1}(\mbf{r}-\mbf{r}_1)\rho_\mathrm{ind,2}(\mbf{r}'-\mbf{r}_2)}{|\mbf{r}-\mbf{r}'|}.\label{eq:edd}\end{equation} With the above formulae, all that remains is to find a relation between the external perturbing charge sources and their induced charge densities via linear response.
To avoid numerical problems with the $r^{-1}$ singularity of the Coulomb potential, and its $k^{-2}$ behavior in reciprocal space, we substitute the external point charges by narrow Gaussians (of 0.25 \AA\ width), what has a negligible effect at the relevant interionic distances.

\begin{figure}[t!]
    \centering
    \includegraphics[scale=0.5]{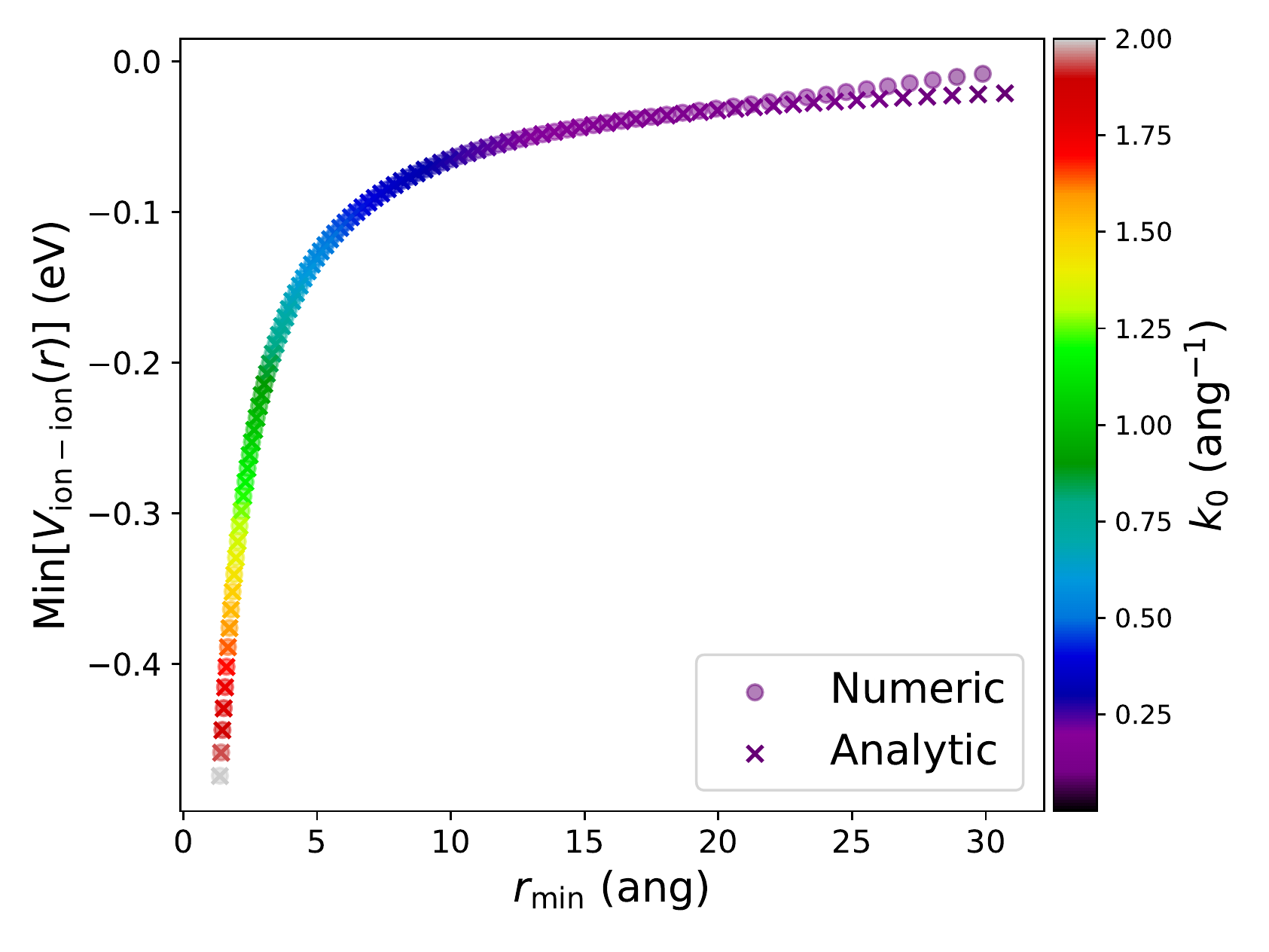}
    \caption{The bound-state depth for dressed-dressed interactions in the TF approximation as a function of its location, colored by the screening parameter $k_0$. The bound state depth gets weaker with increasing screening length, and the position of the minimum shifts to farther separations. The numerical results agree well with the analytical expressions derived from Eq. S3 until small values of $k_0$.}\label{fig:depth-lambda}
\end{figure}

This approximation assumes that each dressing charge is independent of the presence of the other charge, i.e. the response of the medium is linear with the perturbation.
We call this the independent hole approximation (IHA).
We will evaluate the validity of the IHA  calculating (i) the exact interaction of two test charges in a real metal using density functional theory (DFT); and (ii) the exact interaction between two charges in a dielectric liquid (Na$^+$ and Cl$^-$ ions in water).

\begin{figure}[t!]
    \centering\includegraphics[scale=0.5]{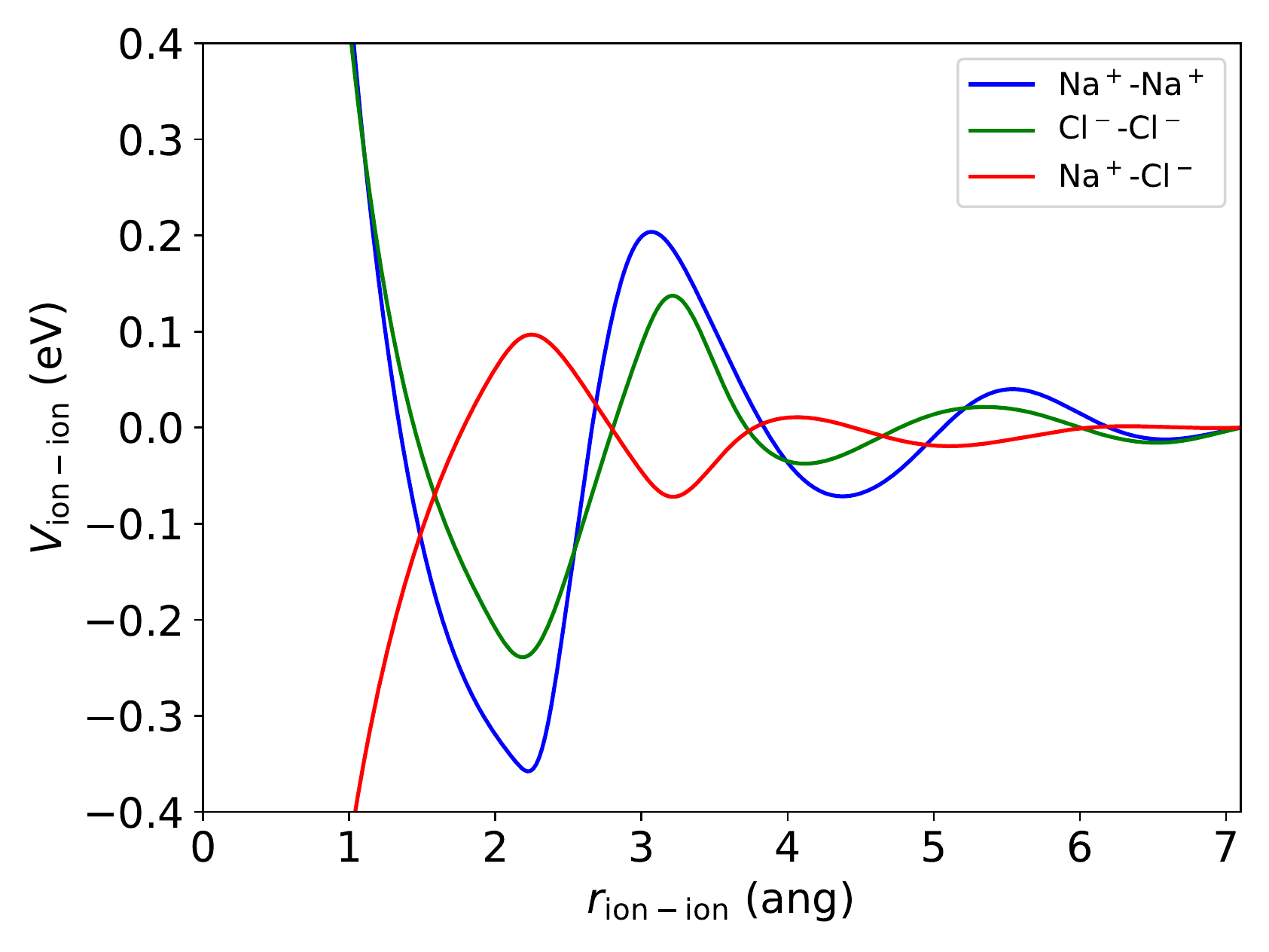}
    \caption{The modeled interaction energy using the induced densities around the ions as defined in Eq.~\eqref{eq:rdfrhoind}. }\label{fig:nnccncrdf}
\end{figure}

\subsection{Model Results}

Fig.~\ref{fig:bdddfigs} shows the model's bare-dressed and dressed-dressed interactions, as well as the induced screening charges given by Eq.~\eqref{eq:rhochiq}.
We use three dielectric functions: the Thomas-Fermi (TF) approximation, the Lindhard or random phase approximation (RPA), and Inkson's interpolating dielectric function with $\epsilon_{k\to0}=80$ \cite{Inkson1971}.
An analytical derivation of the TF interaction potential is provided the supplementary information.
As can be seen for the three approximations, the screened interaction between two like charges is attractive at short distances.
On the other hand, as it should be due to charge inversion, the same interaction will be repulsive for opposite charges. 

The strength (depth of the potential well) of the attractive/repulsive interaction depends on the screening length of the dielectric medium in question.
This can be easily analyzed using the TF model, which allows to control the screening length through the parameter $k_0 = 1/\lambda$. 
The longer this length, the less the charges are screened, and the depth of the potential well goes to zero linearly as $\lambda\rightarrow\infty$.
This behavior is shown in Fig.~\ref{fig:depth-lambda}, where the bound state depth is plotted versus the location of the potential's minimum.
The points are colored by the screening parameter $k_0$.
The stronger the screening, the deeper the bound state becomes.
Conversely, as the screening strength decreases, the bound state depth approaches zero and its position shifts to shorter separations.
Additionally, as the screening strength increases, the minimum shifts to shorter interionic separations, indicative of the stronger localization of the screening charge.

\section{Analysis and Discussion}\label{section:discussion}
\subsection{Charges in Water}

As we have seen, the results presented in Figs.~\ref{fig:spceopls-pmf}, \ref{fig:spceopls-pot}, and \ref{fig:spceopls-coul} agree qualitatively with the model predictions shown in Fig.~\ref{fig:bdddfigs}.
This model assumes that the response of the dielectric medium is linear and that therefore the screening of two charges is the sum of that of the separated charges. 
We can evaluate the validity of this approximation in realistic systems by computing the exact interaction between charges in a self-consistent response of the medium. 
We do this for simulations of a pair of sodium cations and for sodium chloride in liquid water.
Our simulations add to previous ones, since the free energy of solvation of sodium chloride has been well studied \cite{Wills2021,galli_Zhang2020,pmf2_classical_cpmd_Timko2010,pmf1charmmggahfbKhavrutskii2008,pmf3_classical_ihs_qmefp_Ghosh2013,pmf4_classical_scancpmd_Yao2018,pmf5_classical_marcusthry_Roy2017}.

We simulate each lone ion, as well as different pairs of ions, in a box of water.
Using the \textsf{TRAVIS} \cite{travis_Brehm2020} program, we generate the ion-oxygen and ion-hydrogen spatial distribution function (SDF), defined as $g_{i,X}(\mbf{r}) = \langle n_X(\mbf{r})\rangle/n_X$, where $n_X$ is the number density of particle $X$ in the system and $\langle n_X(\mbf{r})\rangle$ is the average number density at a separation $\mbf{r}$ away from the ion.
Then, the charge density of particle $X$ around the ion $i$ is $\rho_{i,X}(\mbf{r}) = Q_X \cdot n_X \cdot g_{i,X}(\mbf{r})$, and
the charge induced around an ion can be constructed as \begin{equation}
    \rho_{i,\mathrm{ind}}(\mbf{r}) = \rho_{i,O}(\mbf{r}) + \rho_{i, H}(\mbf{r}).\label{eq:rdfrhoind} 
\end{equation} 
\begin{figure}[t!]
    \centering
    \includegraphics[scale=0.5]{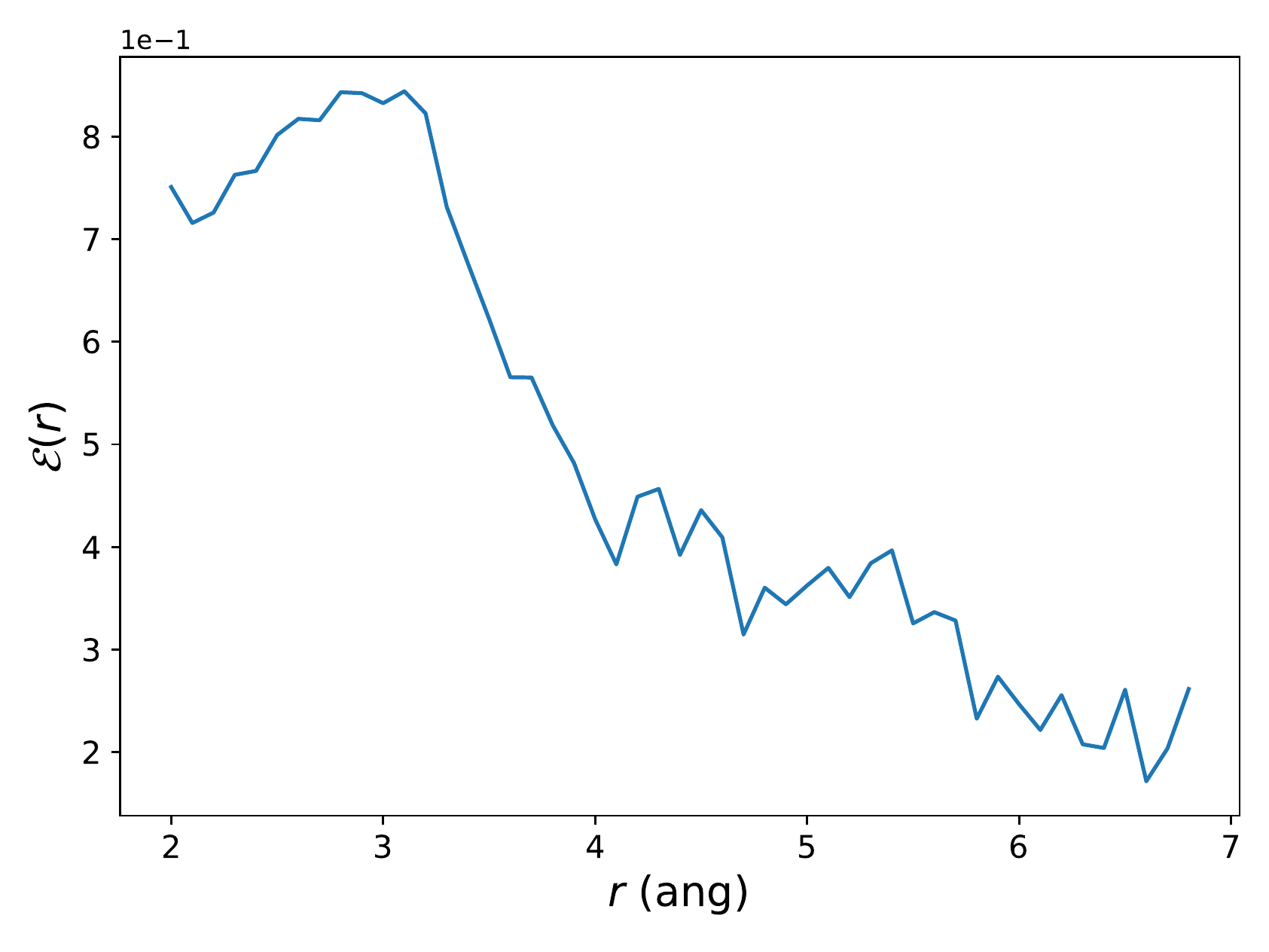}
    \caption{The error as defined in Eq.~\eqref{eq:error} for the densities induced around the \ce{Na+ - Na+} pairs, as a function of their interionic separation $r$. The error increases drastically when the two individual ion SDFs start to overlap.}\label{fig:error-rdf}
\end{figure}

Using this charge in Eq.~\ref{eq:edd}, the dressed-dressed interactions for the \ce{Na+ - Na+}, \ce{Cl- - Cl-}, and \ce{Na+ - Cl-} ion pairs are presented in Fig.~\ref{fig:nnccncrdf}.
Comparison with Figs.~\ref{fig:spceopls-pot} and \ref{fig:spceopls-coul} shows a general qualitative agreement, with similar minima/maxima locations and depths.
There are also some quantitative discrepancies, part of which may be due to the IHA.

\textcolor{black}{
To examine the various effects polarizability, system size, and solution concentration have on our results, we conducted studies to ensure the convergence of these induced charge densities.
Fig. S8 shows the effect that different combinations of non-polarizable \cite{naclt4pew_Joung2008,t4p05Abascal2005} and polarizable \cite{swm4_ions_Yu2010,swm4ndpLamoureux2006} ion and water models have on the modeled dressed-dressed interaction energy.
Here, we see that the polarizability of the ions has little effect on the structure of the surrounding induced charge; the choice of the water model, however, has a large influence in the resulting induced charge density, thus having a greater effect in the modeled interaction.
This is to be expected, as the choice of water model determines the dielectric environment into which the ions are placed.
Figs. S9 and S10 show the results of increasing the bulk system size (i.e., the number of solvent water molecules) and the solution concentration (i.e., the number of solute ions), respectively.
In each case, we find minimal differences in the resulting induced charge structure around the ions, leading to similar predicted dressed-dressed interaction energies.
}

We note, however, that our model considers the ions as point charges, ignoring the Pauli repulsion due to the overlap of the ion cores.
This repulsion is determined by the $\sigma_{ij}$ parameter in the Lennard-Jones potential, which controls at which point the repulsion dominates.
With the OPLS-AA force field, this is 1.8 \AA\ for the sodium ion pair and 2.8 \AA\ for the sodium chloride pair.
It must further be noted that in our model there is no structural information of the solvent besides the approximation for its dielectric function.
In contrast, in the simulations, all this structural information is included in the $g_{i,X}(\mbf{r})$ correlation functions.

We can evaluate the error in the IHA by comparing the exact $\rho_\mathrm{ind}(\mbf{r})$ (calculated from a MD simulation of the ions constrained to be at distance $r$) with the superposition of two copies of the SDF generated via simulation of the lone ion.
This decomposition is shown in Fig. S13 for ion separations of 2.8 and 6.8 \AA.
For each separation constraint $r$, $\rho_\mathrm{ind}(\mbf{r'};r)$ is generated as the cylindrically averaged SDF of the water atoms around the ions.
Then, the SDF from a lone-ion simulation is copied and superposed with itself, with the copies separated by the same separation $r$, giving us $\rho_\mathrm{ind,IHA}(\mbf{r'};r)$.
The difference is taken between $\rho_\mathrm{ind}(\mbf{r'};r)$ and $\rho_\mathrm{ind,IHA}(\mbf{r'};r)$, and the error is found by integrating across the SDF via 
\begin{equation}\mathcal{E}(r)=\frac{\int d\mbf{r'}\cdot\sqrt{\large[\rho_\mathrm{ind}(\mbf{r'};r)-\rho_\mathrm{ind,IHA}(\mbf{r'};r)\large]^2}}{\int d\mbf{r'} \sqrt{\rho_\mathrm{ind}(\mbf{r'};r)^2}}.\label{eq:error}\end{equation}
This error metric for the induced densities around each sodium ion is shown in Fig.~\ref{fig:error-rdf}.
We see that, at $r\sim 3$ \AA\ the error reaches a maximum, arising from the superposed lone-ion SDFs generating charge density in between the two ions, which does not occur in the molecular dynamics simulations with the ion pairs.
The error then steadily decreases as the ions are separated.

 \begin{figure}[t!]
     \centering
     \includegraphics[scale=0.5]{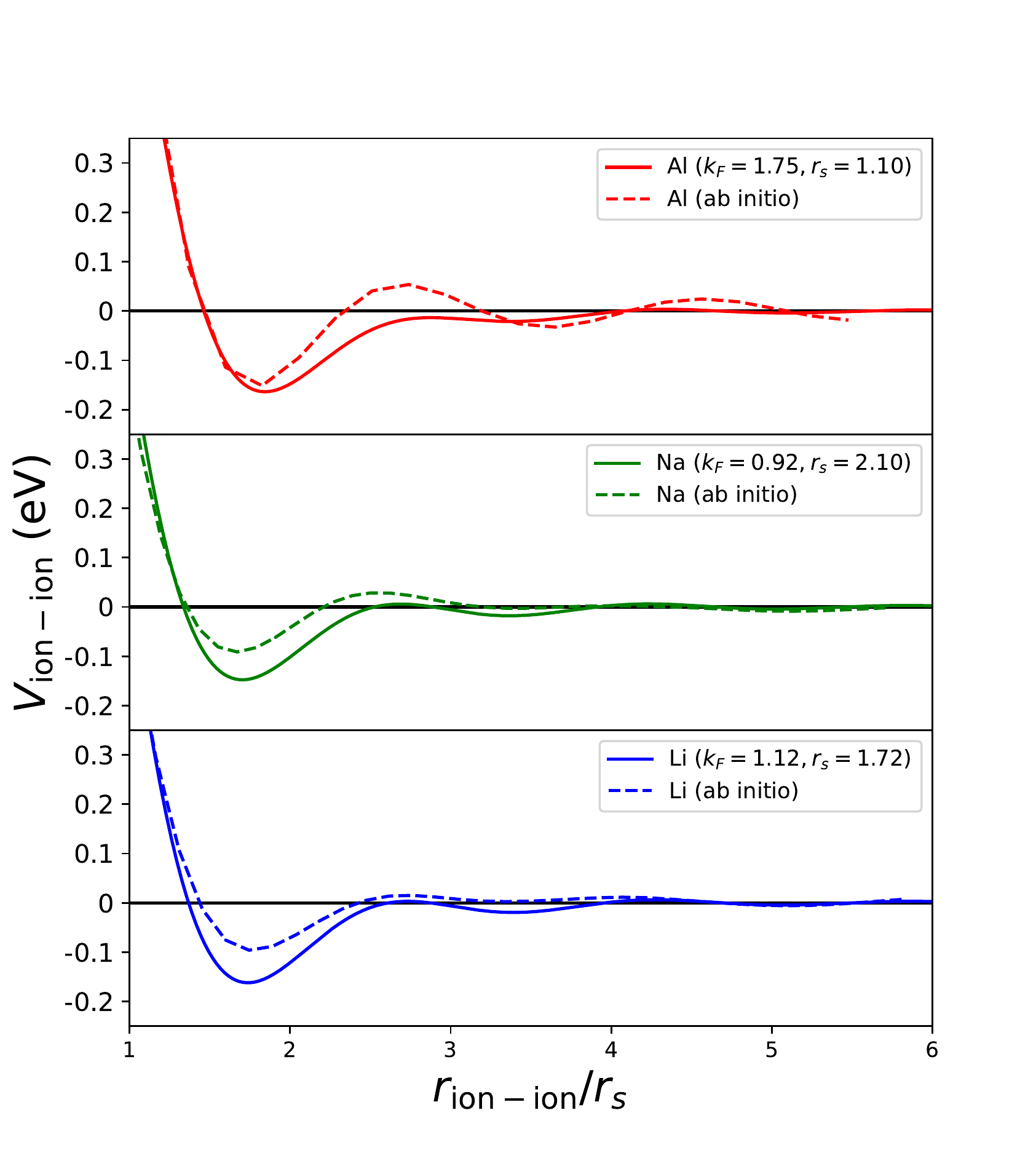}
     \caption{Comparison between our model predictions (solid) and DFT binding curves (dashed) for two hydrogen atoms in aluminum, sodium, and lithium (top to bottom).}
     \label{fig:aimd_model_comp}
\end{figure}

\subsection{Charges in Metals}

Our model is very general and it relies on dielectric functions that are appropriate for the free electron gas.
It is then interesting to compare its predictions with realistic simulations of charges in metals.
The simplest case is a pair of hydrogen atoms, which are fully or partially ionized in metals \cite{Isenberg50}.
The metals can be well-approximated in the RPA regime, whose dielectric function is that of the Thomas-Fermi approximation modified by the Lindhard function $F(q/2k_F)$, with standard Fermi wave vectors for each metal \cite{Ashcroft76}.
Here we examine the proton-proton interactions in aluminum (with a face-centered cubic structure), as well as in sodium and lithium (both body-centered cubic).
We compare the results of our simple model, in the RPA approximation, to those of \textit{ab initio} simulations using density functional theory (DFT).
Details of this simulations are provided in Appendix~\ref{appDFT}.
The model and DFT results are compared in Fig.~\ref{fig:aimd_model_comp} (and in the middle plot of Fig.~\ref{fig:bdddfigs} for $k_F=1$ \AA).
Our model has good qualitative agreement with the \textit{ab initio} results, with the minima locations closely aligned and with the same order of magnitude of depth.
Differences are due to the IHA and to the incomplete ionization of hydrogen atoms, as shown in Fig. S14, where we fit our model to the \textit{ab initio} binding energies using the charge's magnitude and extent as free parameters.
As expected, we find best fits with incompletely but mostly-ionized hydrogen atoms.

\section{Conclusions}

The results above are striking. They show that oppositely charged particles repel, while particles with the same charge attract when embedded in a dielectric with sufficient screening strength, \textit{without} the need for sign inversion of the dielectric function for the medium that causes overscreening.
Instead, it is simply a consequence of the balance of Coulomb interactions between the screening charges and the perturbing sources.
In principle, this behavior is general, and indeed it was already obtained by Corona {\it et al.} for antiparallel-spin electron-electron interaction in a high density homogeneous electron gas \cite{Corona2004}.
We have reproduced their result, finding that the electron-electron interaction has an attractive well, although too weak and short-ranged to produce a bound state of the electron pairs.

In this work, we have shown that a simple electrostatic model
based on linear response reproduces very nicely the behaviour of classical charges
in a dielectric medium.
Using this model, we can explain the origin of the various solvation states in the potential of mean force between ions in solution.
The decomposition of the PMF further illuminates the role of entropy in the solvate's attraction and repulsion.

Even more, this simple model nicely reproduces the pairing of protons in metals
as obtained from \textit{ab initio} DFT simulations.
It is tempting at this point to make connections with electron pairing mechanisms that lead to superconductivity, such as Cooper pairs.
As explained above, the observation of an attractive dressed-dressed electron-electron interaction is not new, although the attractive region of the potential was not highlighted \cite{Corona2004} and an attractive well does not imply the existence of a bound quantum state. 
However, what this work has shown is that it is possible to achieve an attractive effective interaction and that this interaction will depend only on the spatial distribution of the electronic exchange and correlation hole.

Independently of its possible quantitative importance in a variety of systems, we have shown a general and counter-intuitive physical effect that is not generally recognized.
There is no need to invoke overscreening to explain an attractive interaction between like charges \cite{Allen1988_totaldiel,Kornyshev1997_overscreening}, and hence this should be taken into account when asserting the structural and dynamical properties of ions and charges in solution or at interfaces, including at the electrochemical interface.
This also highlights the necessity to have very accurate charged models for the solvent, given that the structural properties resulting from the model parameters will have a profound impact on the behavior of the screening.

\section*{Acknowledgements}

We would like to thank the Institute for Advanced Computational Science (IACS) for the use of their SeaWulf and Ookami high performance computing clusters, where our simulations took place. This work was in part funded by the U.S. Department of Energy, Office of Science, Basic Energy Sciences, under Award No. DE-SC0019394, as part of the CCS Program, by Spain's Ministry of Science grant PID2022-139776NB-C64, and by the Graduate Fellowship program offered at the IACS.

\appendix

\section{Ion-Ion Screening in Metals}\label{appDFT}

\textit{Ab initio} energy calculations were done using the density functional theory (DFT) code \textsf{SIESTA} with a van der Waals exchange-correlation functional (vdW-DRSLL).
Pseudopotentials were used for all core electrons and a triple-zeta polarized (TZP) basis set was used for all valence electrons.

Two H atoms  were added to three different $5\times5\times5$ metallic supercells. The aluminum lattice is fcc with a unit cell lattice constant of 4.05 \AA. The sodium and lithium lattices are bcc with unit cell lattice constants of 4.29 \AA\ and 3.51 \AA\, respectively.
The placement of the atoms was made to ensure that they would not come into contact with any aluminum nuclei.
The plots shown in Fig. S1
are obtained after convoluting the H-H distance along the [001] direction, in order to remove the influence of crystal potential
on the results.
The two atoms were separated by distances between 0.25 \AA\ and 6$r_s$ at intervals of 0.25 \AA\, while maintaining the same midpoint position.
The energies were adjusted so that 0 eV corresponds to the asymptote for large atomic separations.

\bibliography{__refs}

\end{document}


\preprint{AIP/123-QED}

\title[]{Supplemental: Anti-Coulomb ion-ion interactions: a theoretical and computational study}

\newcommand{\SBUphysast}{Physics and Astronomy Department, Stony Brook University, Stony Brook, New York 11794-3800, United States}
\newcommand{\IACS}{Institute for Advanced Computational Science, Stony Brook University, Stony Brook, New York 11794-3800, United States}
\newcommand{\UADM}{Departamento e Instituto de Física de la Materia Condensada (IFIMAC), Universidad Aut\'{o}noma de Madrid, E-28049 Madrid, Spain}

\author{Alec Wills}
\email{alec.wills@stonybrook.edu}
\affiliation{\SBUphysast}
\affiliation{\IACS}
\author{Anthony Mannino}
\affiliation{\SBUphysast}
\affiliation{\IACS}
\author{Isidro Losada}
\affiliation{\UADM}
\author{Sara G. Mayo}
\affiliation{\UADM}
\author{Jose M. Soler}
\affiliation{\UADM}
\author{Marivi Fern\'{a}ndez-Serra}%
 \email{maria.fernandez-serra@stonybrook.edu}
\affiliation{\SBUphysast}
\affiliation{\IACS}

\date{\today}

\begin{abstract}


\end{abstract}

\maketitle

\section{Dressed-Dressed Interaction Potential in the Thomas-Fermi Approximation:}\label{app:TF}

We show here the analytical derivation for dressed-dressed interaction in the simplest case: the Thomas-Fermi model \cite{Ashcroft76}.
In this case, the  screened potential of a charge Q in a metal is $\Phi(r)=\frac{Q}{r}e^{-k_0r}$. The associated screening/dressing charge is:

\begin{align*}
\rho(r) &= -\frac{1}{4\pi}\nabla^2\phi(r)\\
&=-\frac{1}{4\pi}\frac{1}{r}\frac{d^2}{dr^2}r\phi(r)\\
&=-\frac{Q}{4\pi r}k_0^2e^{-k_0r}\\
&=-\frac{k_0^2}{4\pi}\phi(r).\numberthis
\end{align*}

Its Fourier transform is given by

\begin{align*}
\rho(\mbf{q}) &= \frac{1}{(2\pi)^{3/2}}\int{d^3r\rho(\mbf{r})e^{-i\mbf{q}\cdot\mbf{r}}}\\
&= \frac{1}{(2\pi)^{3/2}}\int_0^\infty{4\pi r^2dr j_0(qr)\rho(r)}.\numberthis
\end{align*}
%
Given that the expansion of a plane wave in spherical harmonics is given by $e^{iqr}=\sum_{l,m}i^lj_0(qr)Y_{lm}(\hat{\mbf{q}})Y_{lm}(\hat{\mbf{r}})$, we have:

\begin{align*}
\rho(q) &= \frac{4\pi}{(2\pi)^{3/2}}\int_0^\infty{r^2dr\frac{sin(qr)}{qr}\frac{-k_0^2Q}{4\pi r}e^{-k_0r}}\\
&=-\frac{Qk_0^2}{(2\pi)^{3/2}(k_0^2+q^2)}\numberthis
\end{align*}

Then the interaction between two charge densities is:

\begin{widetext}
\begin{align*}
V_{12}&= \int{d^3r \phi_1(\mbf{r}-\mbf{r_1})\rho_2(\mbf{r}-\mbf{r_2})} \\
&=\int d^3r \left[\frac{1}{(2\pi)^{3/2}}\int d^3q_1 \phi_1(\mbf{q_1}) e^{i \mbf{q_1}\cdot(\mbf{r}-\mbf{r}_1)} \right] \left[\frac{1}{(2\pi)^{3/2}}\int d^3q_2\rho_2(\mbf{q_2}) e^{i \mbf{q_2}\cdot(\mbf{r}-\mbf{r}_2)} \right] \\
&= \frac{1}{(2\pi)^3}\int d^3q_1d^3q_2 \phi_1(\mbf{q_1})\rho_2(\mbf{q_2})e^{-i (\mbf{q_1}\cdot\mbf{r_1}+\mbf{q_2}\cdot\mbf{r_2})}\int d^3r e^{i\mbf{r}\cdot(\mbf{q_1}+\mbf{q_2})}\\
&= \int d^3q \phi_1(-\mbf{q})\rho_2(\mbf{q})e^{-i\mbf{q}\cdot(\mbf{r_2}-\mbf{r_1})}\\
&=4\pi\int q^2dq\left[\frac{4\pi}{q^2}\rho_1(q)\right]\rho_2(q)\frac{sin(qr_{12})}{qr_{12}}\\
&=16\pi^2\int{dq \rho_1(q)\rho_2(q)\frac{sin(qr_{12})}{qr_{12}}}\\
&=Q_1Q_2k_{0,1}^2k_{0,2}^2\cdot \frac{2}{\pi}\int_0^\infty {dq \frac{1}{(k_{0,1}^2+q^2)(k_{0,2}^2+q^2)}\frac{sin(qr_{12})}{qr_{12}}},\numberthis
\end{align*}
\end{widetext}
where $q=|\mbf{q}|$ and $r_{12}=|\mbf{r}_2-\mbf{r}_1|$.
The integral over $d^3r$ enforces $\mbf{q_1}=-\mbf{q_2}$, but since $\rho$ is even in $q$ the negative signs are dropped upon substituting $\rho(q)$ for $\phi(q)$.

For two charges in the same medium the screening decay parameters $k_0$ are identical, so \begin{equation}
\begin{split}
V_{12}^{scr}=Q_1Q_2 k_0^4\cdot \frac{2}{\pi} \int_0^\infty \frac{dq}{(k_0^2+q^2)^2} \frac{sin(qr_{12})}{qr_{12}},
\end{split}
\end{equation} and the total interaction between the screened charges is: \begin{equation}
\begin{split}
V_{12}^{tot} = -\frac{Q_1Q_2}{r_{12}}+\frac{2Q_1Q_2}{r_{12}} e^{-k_0r_{12}}+V_{12}^{scr},
\end{split}
\end{equation} where the second term is each point charge's interaction with the other's induced charge density and the third term is the interaction between the induced charges.
The first term is the bare-bare interaction of the point charges, and is negative to subtract out the double counting of this interaction included in the bare-dressed energy.
This is necessary since the analytical forms of $\rho_\mathrm{TF}$ and $\phi_\mathrm{TF}$ are the \textit{total induced} density and potential, including the point charge that is screened.
%


Defining $x\equiv k_0r_{12}$ and $\Tilde{{q}}\equiv q/k_0$, \begin{equation}
\begin{split}
V_{12}^{tot} &= Q_1Q_2k_0\left[  -\frac{1}{x}+\frac{2}{x}e^{-x}   +\frac{2}{\pi}\int_0^\infty \frac{d\Tilde{q}}{(1+\Tilde{q}^2)^2} \frac{sin(\Tilde{q}x)}{\Tilde{q}x} \right]\\
&=\frac{Q_1Q_2k_0}{x}\left[-1+2e^{-x} + \frac{2}{\pi} \int_0^\infty \frac{dy}{\left(1+\left[\frac{y}{x}\right]^2\right)^2} \frac{sin(y)}{y}\right],\label{supp-eq:vtot12}
 \end{split}
\end{equation} where for the final equality we let $y=\Tilde{q}x$.
This potential has a  minimum of $V_\mathrm{min}=-8.72\times 10^{-3}\cdot\frac{Q_1Q_2}{r_0}$ H at $r=2.73r_0$, with $r_0=1/k_0=\frac{\sqrt{r_s}}{2}(2\pi/3)^{1/3}$.
These $(r_\mathrm{min},V_\mathrm{min})$ values as a function of $k_0$ are plotted alongside numerically derived results in Fig. 5.
The agreement between the linear response numerical results and the analytically derived expressions is exact until departures begin in the regime of small $k_0$.

\section{Phenomenological Model for Ion-Ion Screening in Metals}
\begin{figure}[h!]
\centering
    \includegraphics[scale=0.5]{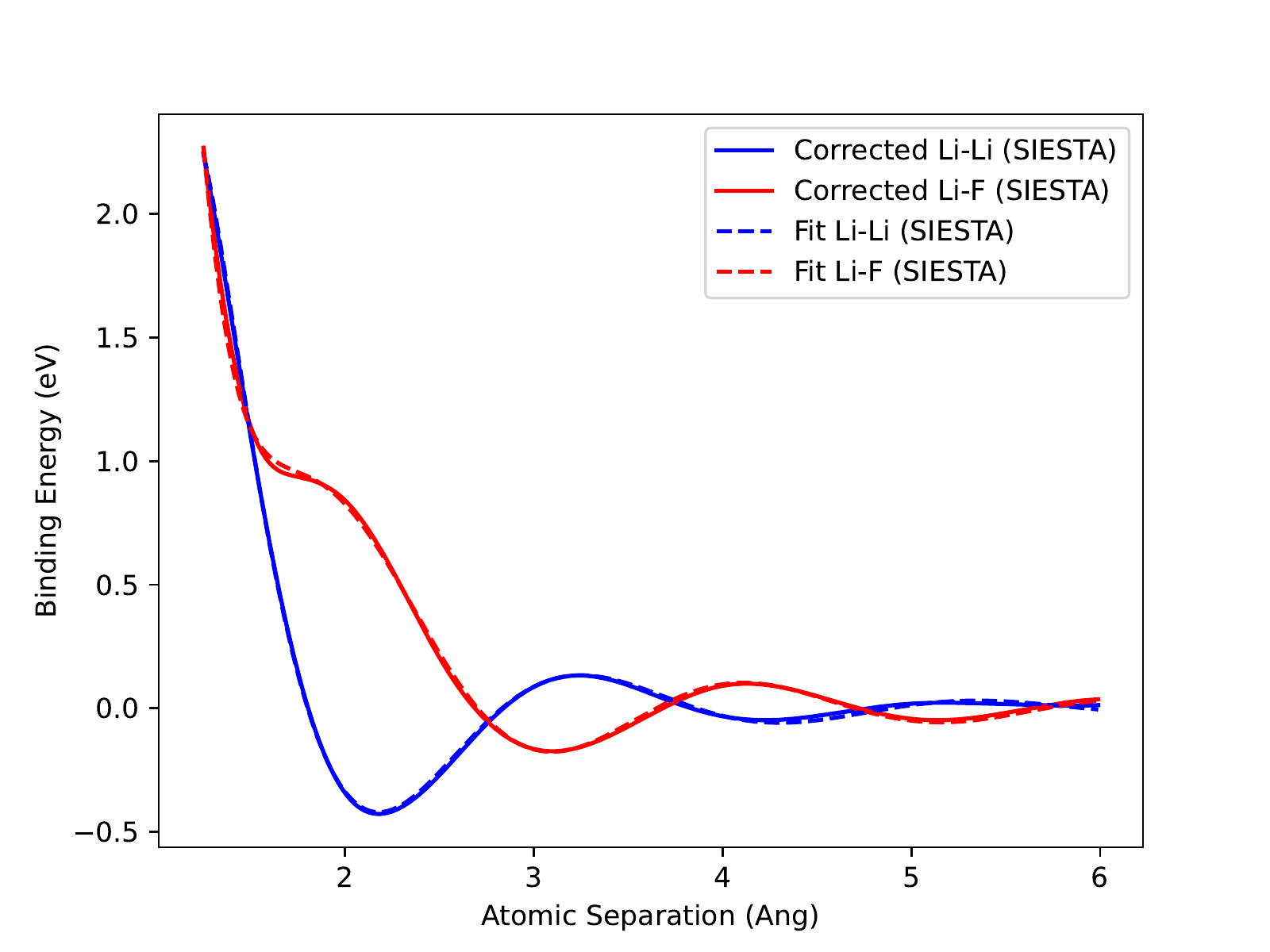}
        \caption{Total energy calculations (corrected for bare ion interactions) for Li-Li (blue) and Li-F (red) with the aluminum lattice. The corresponding dashed lines are fits using \eqref{eq:fit}.}
        \label{supp-fig:corr_SIESTA}
\end{figure}

Our energy calculations within the aluminum lattice show the full interaction between the two ions in the metal.
It is useful to break this total interaction down into constituent parts that can be explained physically.
The first correction we make is for the bare-bare ionic interaction.
This can be described as a simple Coulomb potential between two point charges.
Mulliken charges for each atom in the system are produced in the \textsf{SIESTA} output file.
The Coulomb interaction between the two relevant ions in the system is subtracted from each energy calculation.
These results are shown in Fig.~\ref{supp-fig:corr_SIESTA}.
This figure shows much more symmetric energy energy curves with nearly identical periodicity.
We fitted these curves to the following functional form.
\begin{equation}
    E = a+b\frac{cos(cx+d)}{x^e}+f\Bigg[\bigg(\frac{g}{x}\bigg)^{12}-\bigg(\frac{g}{x}\bigg)^6\Bigg] \label{eq:fit}
\end{equation}
Here, E is the energy, x is the ionic separation distance, and a-e are parameters of the fit.
The first term is for energy scaling, the second term is inspired by the Lindhard model (where c should be proportional to the Fermi wavevector), and the third term has the form of the Lennard-Jones potential.
We see that the fit is very strong and we obtain a consistent value for c in each fit.

\bibliography{__refs}

\begin{figure}[h!]
    \centering
    \includegraphics[scale=0.5]{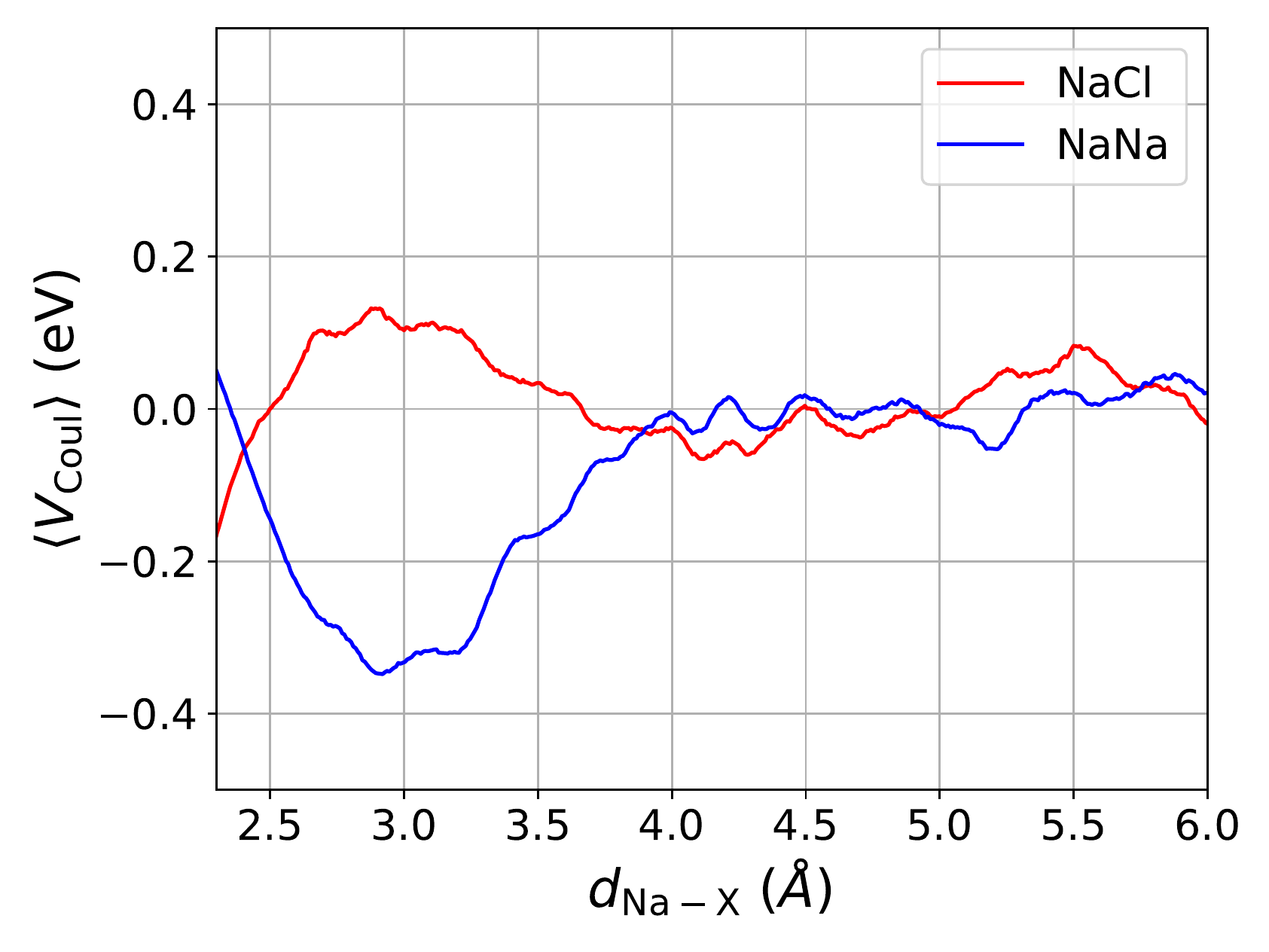}
    \caption{The average Coulomb energy as a function of interionic distance with JCP ion parameters in SPC/E water. The transition barrier between oppositely charged ions (NaCl, red) is clearly visible as the ions are brought together. On the other hand, there is a strong attraction between the two like-charged ions (NaNa, blue).}\label{supp-fig:spcejcp-coul}
\end{figure}

\begin{figure*}[!ht]
     \centering
     \begin{subfigure}[b]{0.32\textwidth}
     \includegraphics[scale=0.32]{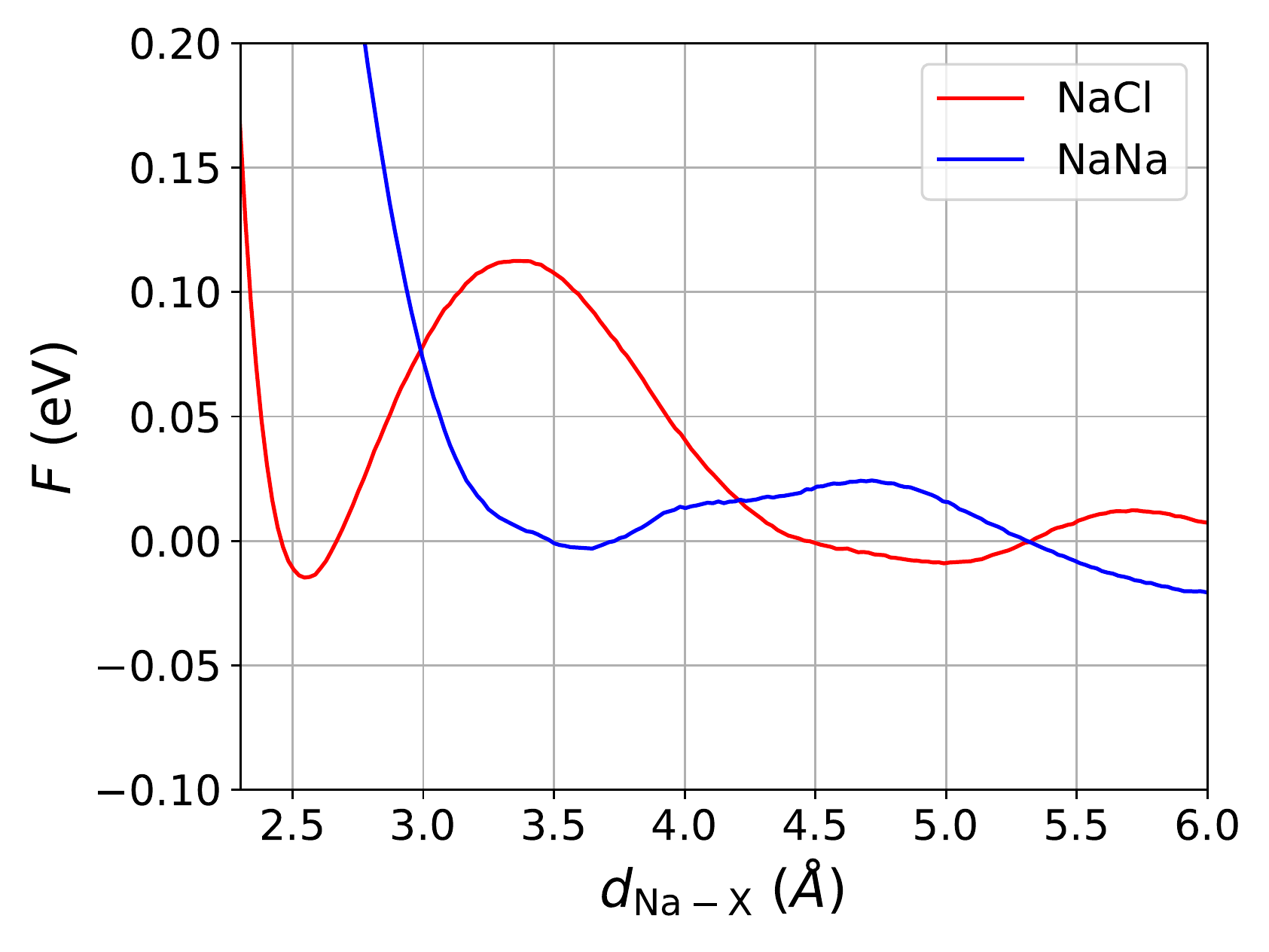}
         \caption{}
         \label{supp-fig:spcejcp-pmf}
     \end{subfigure}
     \begin{subfigure}[b]{0.32\textwidth}
     \includegraphics[scale=0.32]{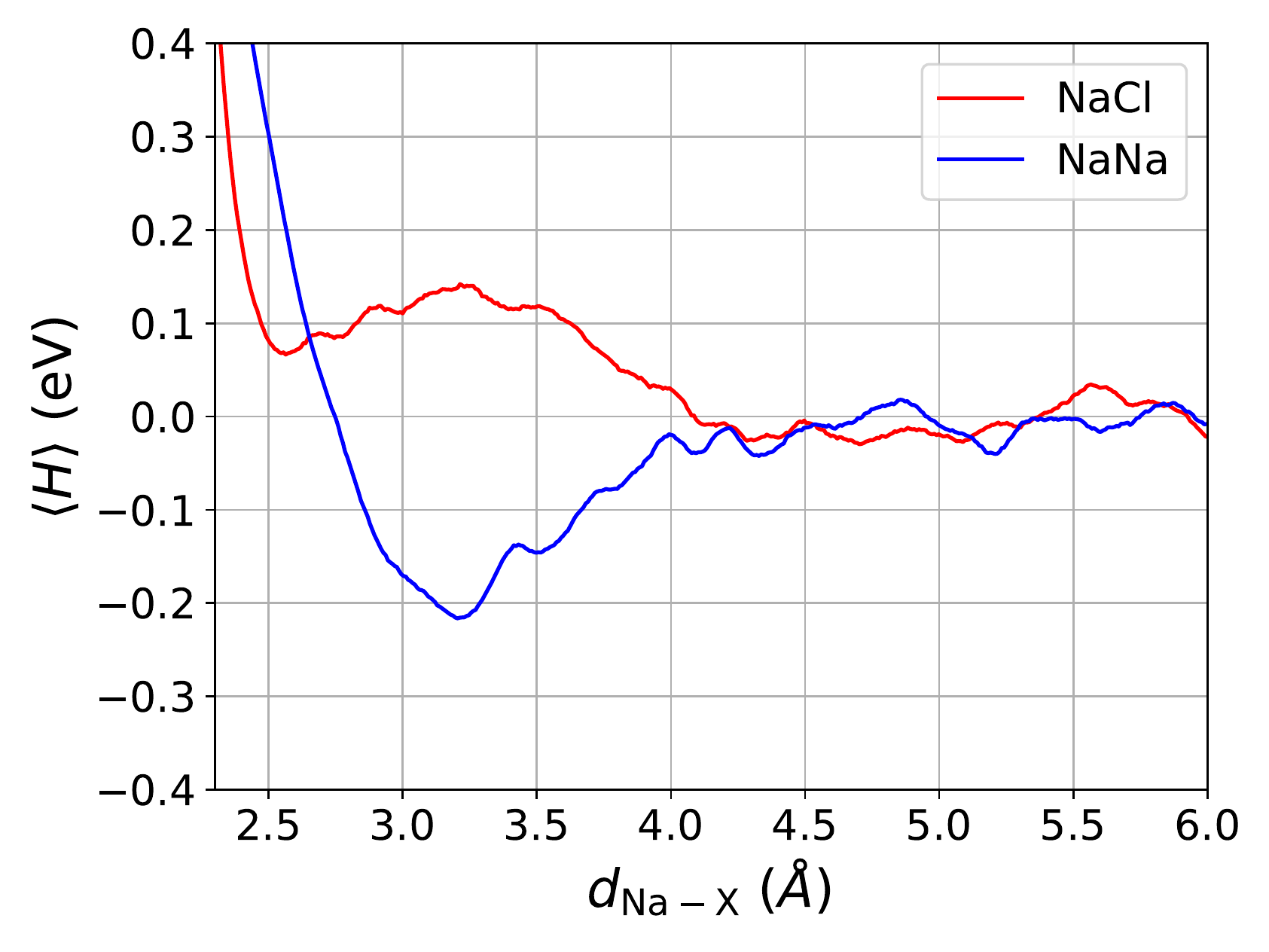}
         \caption{}
         \label{supp-fig:spcejcp-pot}
     \end{subfigure}
     ~ 
     \begin{subfigure}[b]{0.32\textwidth}
     \includegraphics[scale=0.32]{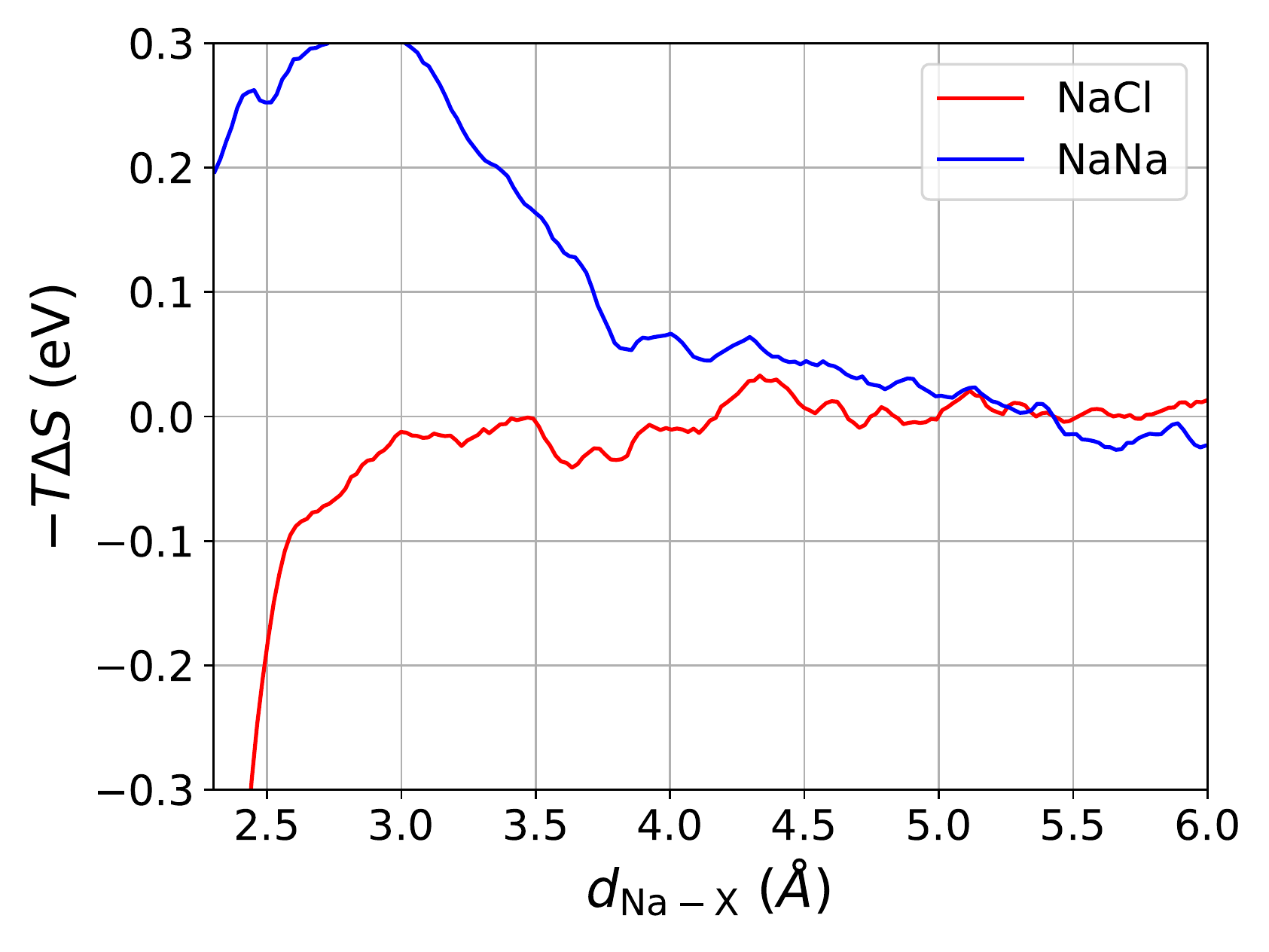}
         \caption{}
         \label{supp-fig:spcejcp-tds}
     \end{subfigure}
     \caption{\textbf{(a)} The WHAM generated NVT PMFs for JCP NaCl (red) and NaNa (blue) ions solvated in SPC/E the water model. \textbf{(b)} The histogram-binned total potential energies along the reaction coordinate, giving the enthalpic contribution to the free energy. \textbf{(c)} The entropic contribution to the free energy, taken as $-T\Delta S = F - H$.}
\end{figure*}

\begin{figure}
    \centering
    \includegraphics[scale=0.5]{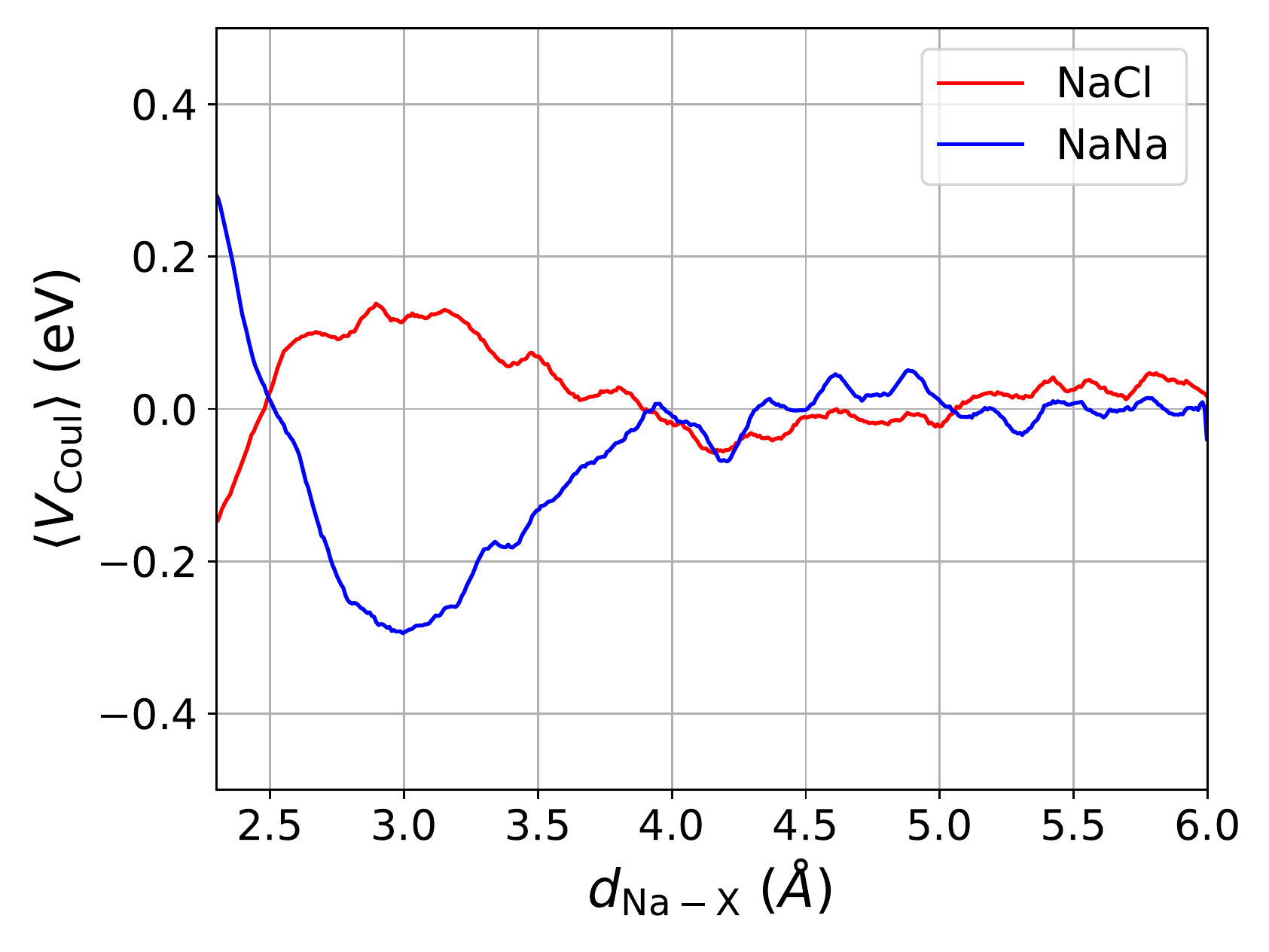}
    \caption{The average Coulomb energy as a function of interionic distance for OPLS-AA ion parameters in TIP4P water. The transition barrier between oppositely charged ions (NaCl, red) is clearly visible as the ions are brought together. On the other hand, there is a strong attraction between the two like-charged ions (NaNa, blue).}\label{supp-fig:t4popls-coul}
\end{figure}

\begin{figure*}[!ht]
     \centering
     \begin{subfigure}[b]{0.32\textwidth}
     \includegraphics[scale=0.32]{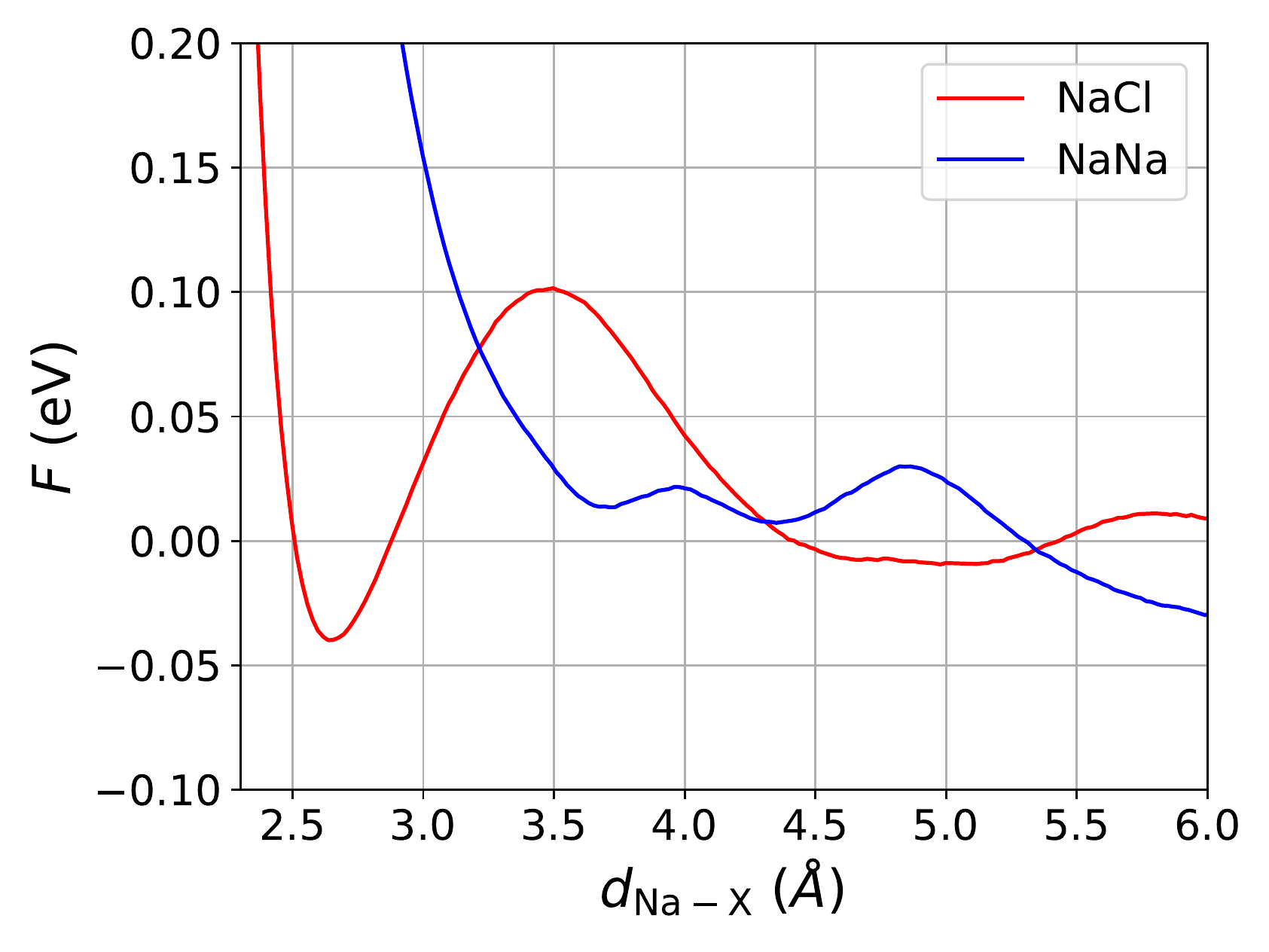}
         \caption{}
         \label{supp-fig:t4popls-pmf}
     \end{subfigure}
     \begin{subfigure}[b]{0.32\textwidth}
     \includegraphics[scale=0.32]{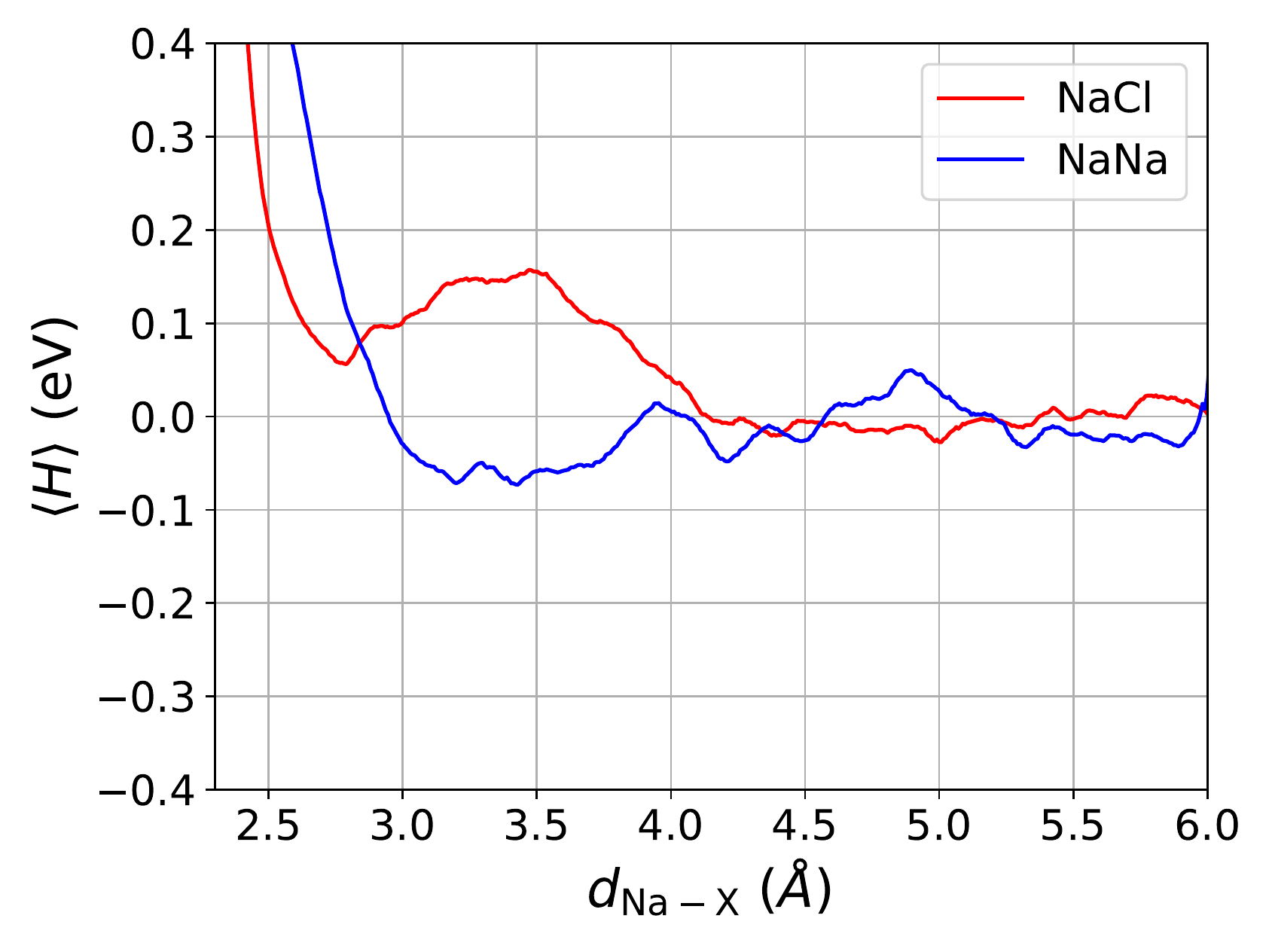}
         \caption{}
         \label{supp-fig:t4popls-pot}
     \end{subfigure}
     ~ 
     \begin{subfigure}[b]{0.32\textwidth}
     \includegraphics[scale=0.32]{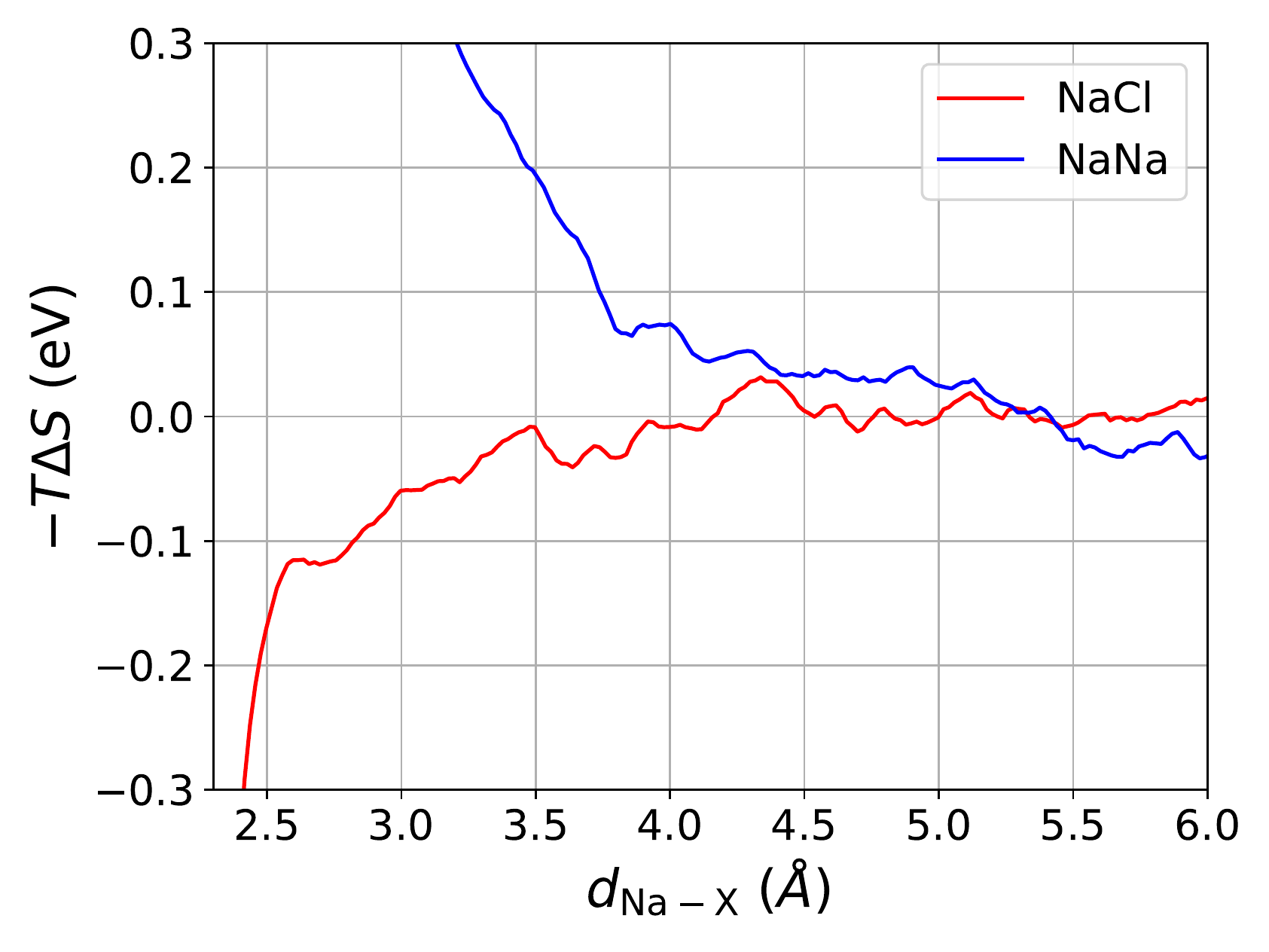}
         \caption{}
         \label{supp-fig:t4popls-tds}
     \end{subfigure}
     \caption{\textbf{(a)} The WHAM generated NVT PMFs for OPLS-AA NaCl (red) and NaNa (blue) ions solvated in TIP4P the water model. \textbf{(b)} The histogram-binned total potential energies along the reaction coordinate, giving the enthalpic contribution to the free energy. \textbf{(c)} The entropic contribution to the free energy, taken as $-T\Delta S = F - H$.}
\end{figure*}

\begin{figure}
    \centering
    \includegraphics[scale=0.5]{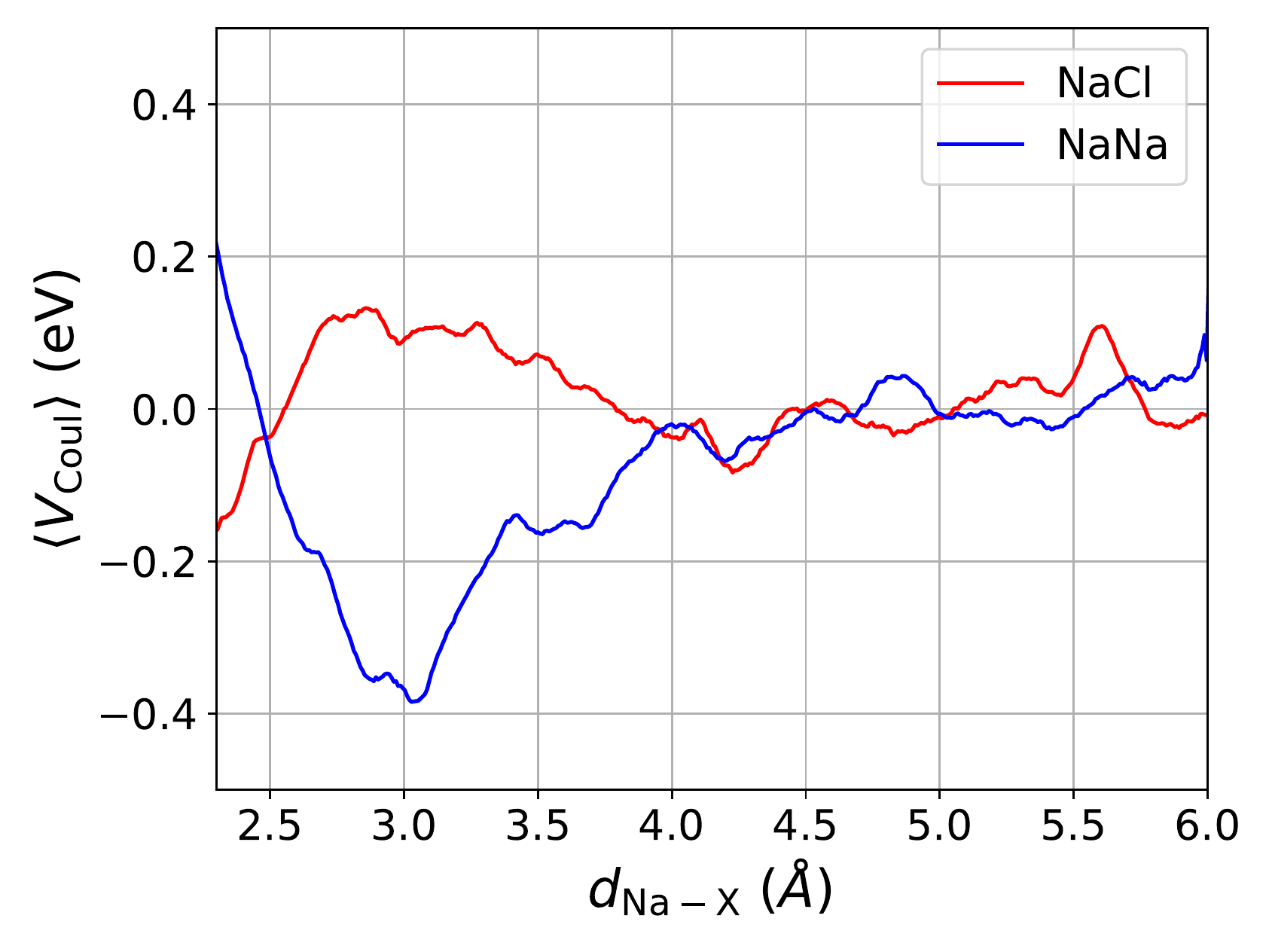}
    \caption{The average Coulomb energy as a function of interionic distance for OPLS-AA ions in TIP4P/Ew water. The transition barrier between oppositely charged ions (NaCl, red) is clearly visible as the ions are brought together. On the other hand, there is a strong attraction between the two like-charged ions (NaNa, blue).}\label{supp-fig:t4pewopls-coul}
\end{figure}

\begin{figure*}[!ht]
     \centering
     \begin{subfigure}[b]{0.32\textwidth}
     \includegraphics[scale=0.32]{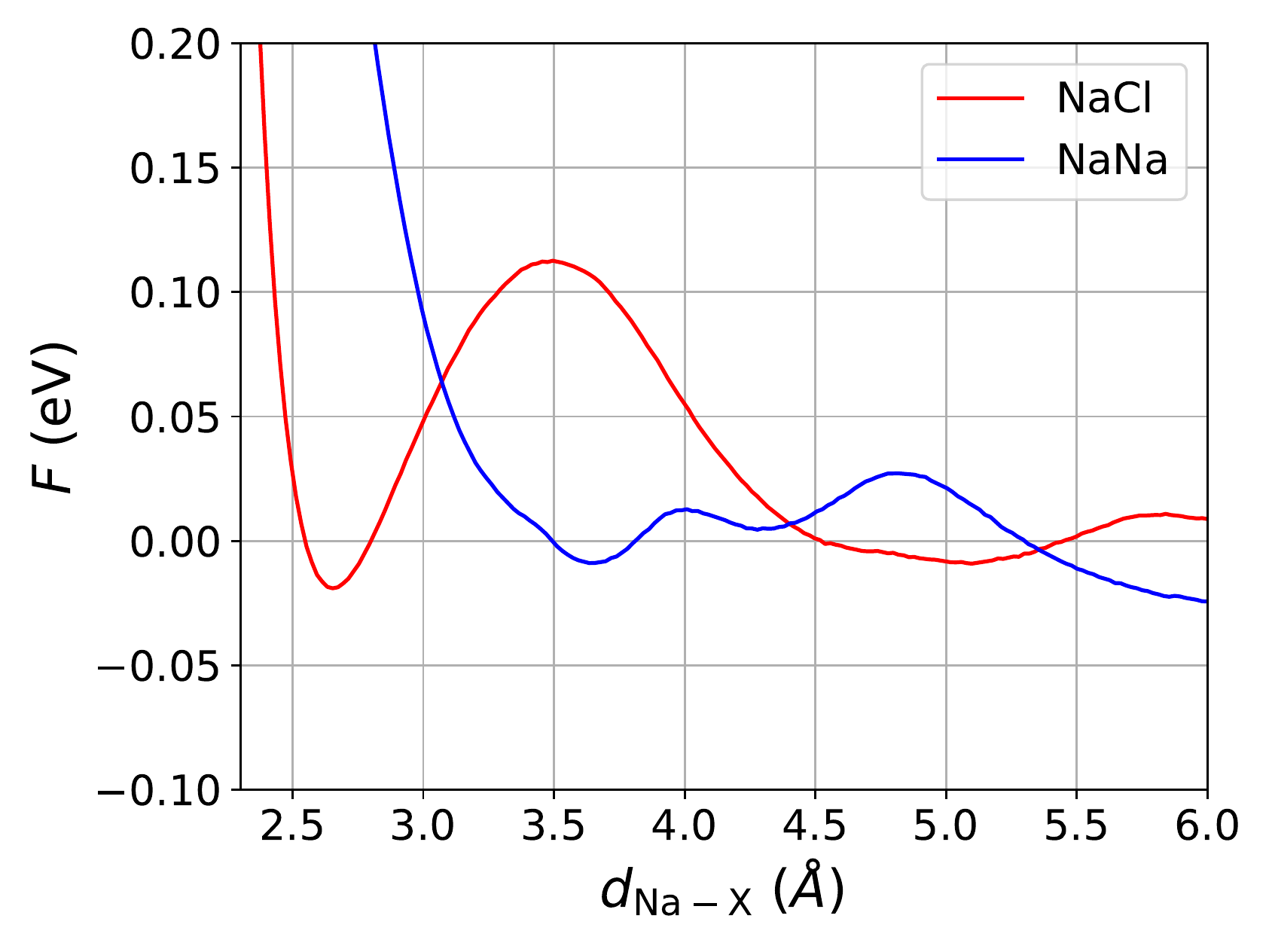}
         \caption{}
         \label{supp-fig:t4pewopls-pmf}
     \end{subfigure}
     \begin{subfigure}[b]{0.32\textwidth}
     \includegraphics[scale=0.32]{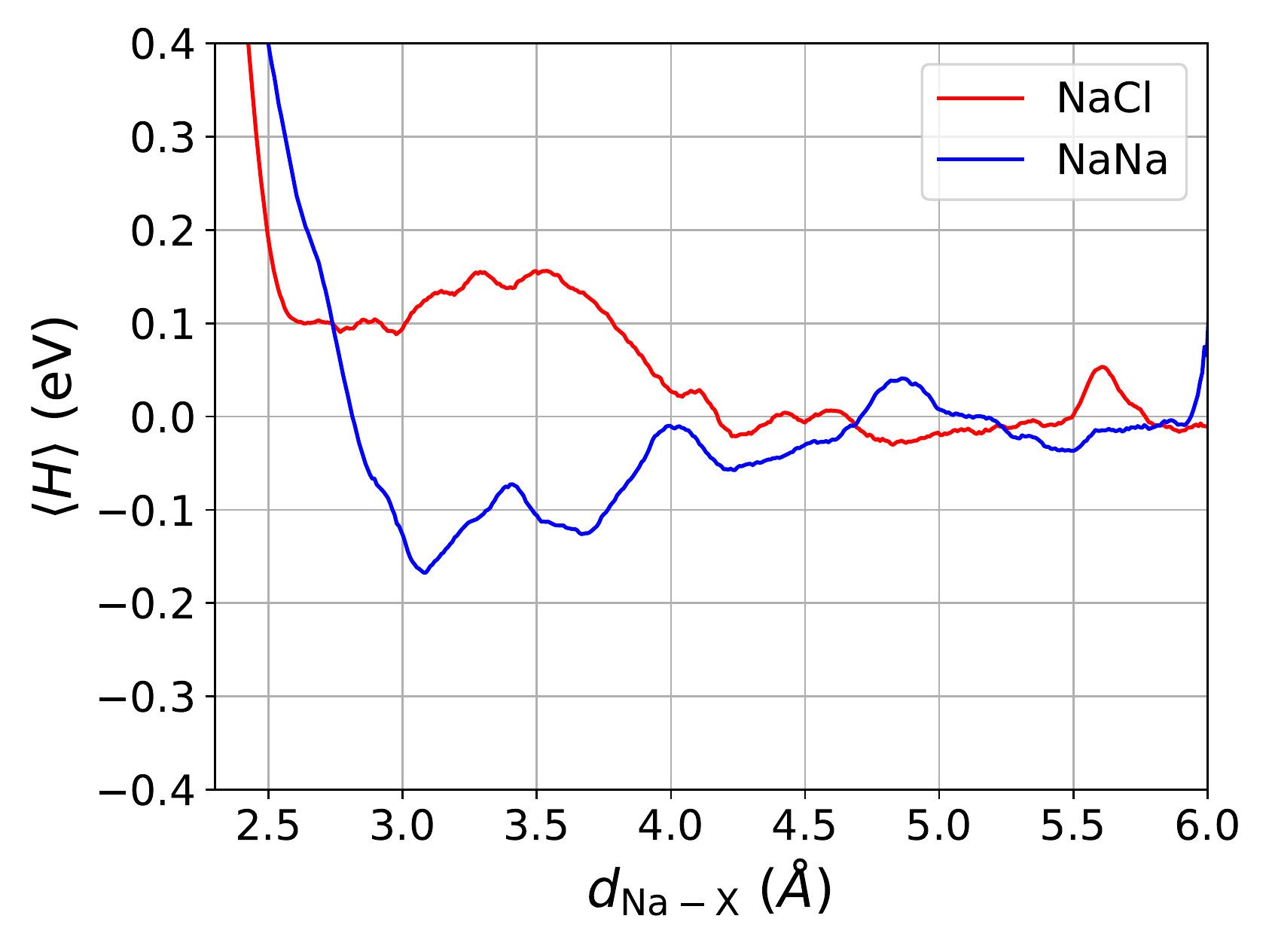}
         \caption{}
         \label{supp-fig:t4pewopls-pot}
     \end{subfigure}
     ~ 
     \begin{subfigure}[b]{0.32\textwidth}
     \includegraphics[scale=0.32]{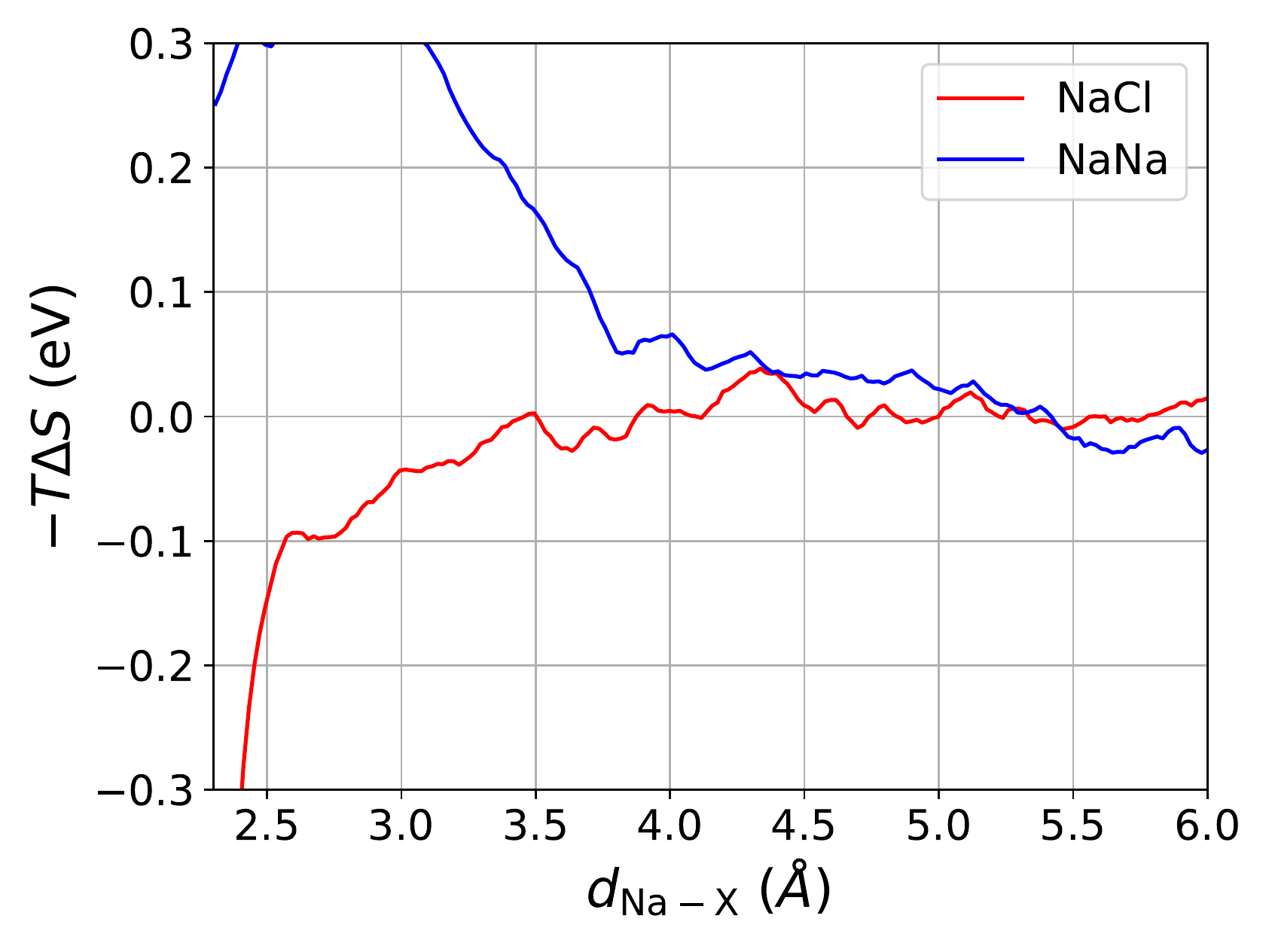}
         \caption{}
         \label{supp-fig:t4pewopls-tds}
     \end{subfigure}
     \caption{\textbf{(a)} The WHAM generated NVT PMFs for OPLS-AA NaCl (red) and NaNa (blue) ions solvated in TIP4P/Ew the water model. \textbf{(b)} The histogram-binned total potential energies along the reaction coordinate, giving the enthalpic contribution to the free energy. \textbf{(c)} The entropic contribution to the free energy, taken as $-T\Delta S = F - H$.}
\end{figure*}

\begin{figure}
	\centering
	\includegraphics[scale=0.5]{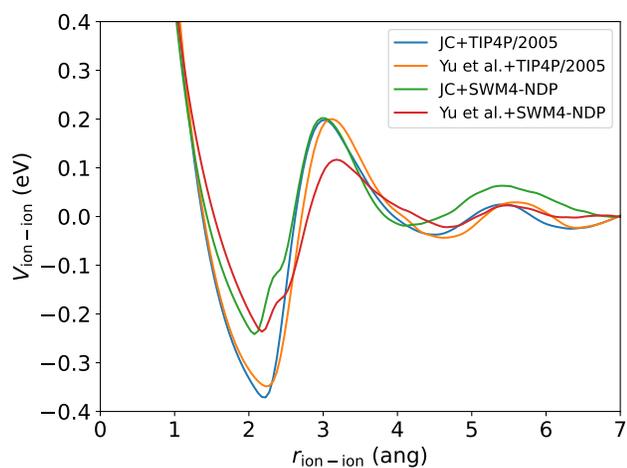}\\
	\caption{Modeled dressed-dressed interaction energies between ion pairs, with varying combinations of non-polarizable \cite{naclt4pew_Joung2008} and polarizable \cite{swm4_ions_Yu2010} ion parameters and the TIP4P/2005 non-polarizable \cite{t4p05Abascal2005} and SWM4-NDP polarizable water models \cite{swm4ndpLamoureux2006}. Results indicate the polarizability of the ions has little effect on the resulting predicted interaction energy. Rather, the dielectric properties of the solvent that determine the resulting structure have a more important effect.}\label{supp-fig:swm4vdd_pols}
\end{figure}

\begin{figure}
	\centering
	\includegraphics[scale=0.5]{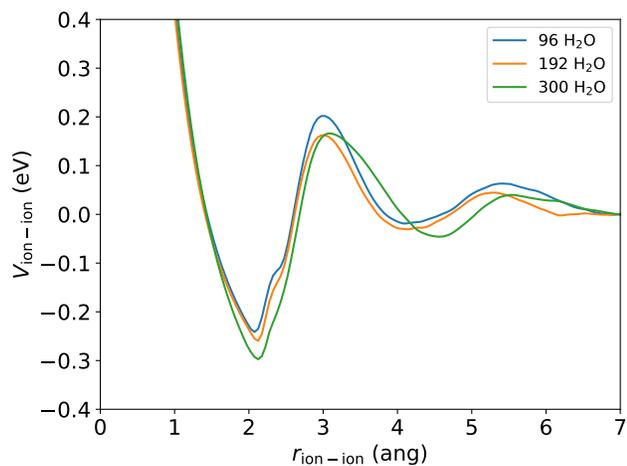}\\
	\caption{Modeled dressed-dressed interaction energies with varying sizes and amount of bulk water. Our results show no indication that finite-size effects affect our predicted energies in our simulated systems.}\label{supp-fig:swm4vdd_boxes}
\end{figure}
\begin{figure}
	\centering
	\includegraphics[scale=0.5]{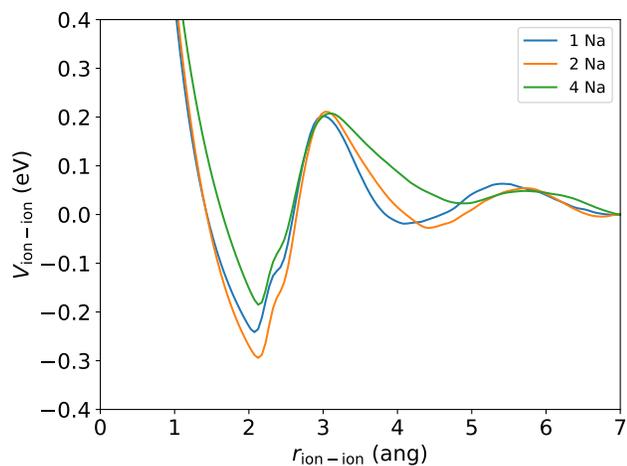}\\
	\caption{Modeled dressed-dressed interaction energies between ions in a box of 96 water molecules, with varying numbers of solute ions. The results indicate that, while the presence of other ions may interfere with the resulting induced charge structure, the overall behavior remains as our model predicts.}\label{supp-fig:swm4vdd_ions}
\end{figure}
\begin{figure}
	\centering
	\includegraphics[scale=0.5]{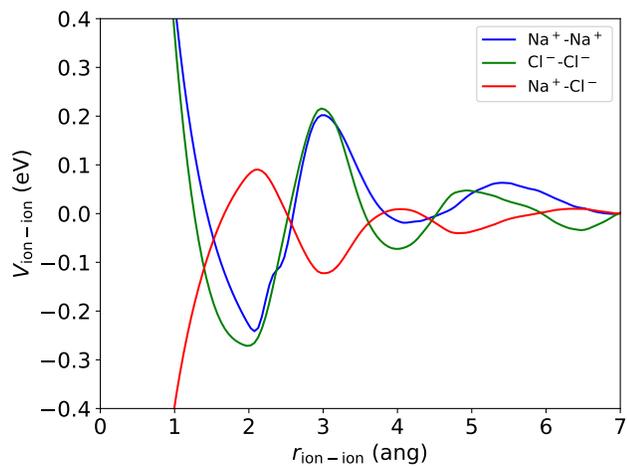}\\
	\caption{Modeled dressed-dressed interaction energies between ion pairs\cite{naclt4pew_Joung2008} in the polarizable SWM4-NDP\cite{swm4ndpLamoureux2006} water. Results are qualitatively similar to those presented in the main body, although the bound-state minima are slightly less deep than in Fig. 6.}\label{supp-fig:swm4vdd}
\end{figure}

\onecolumngrid

\begin{figure}
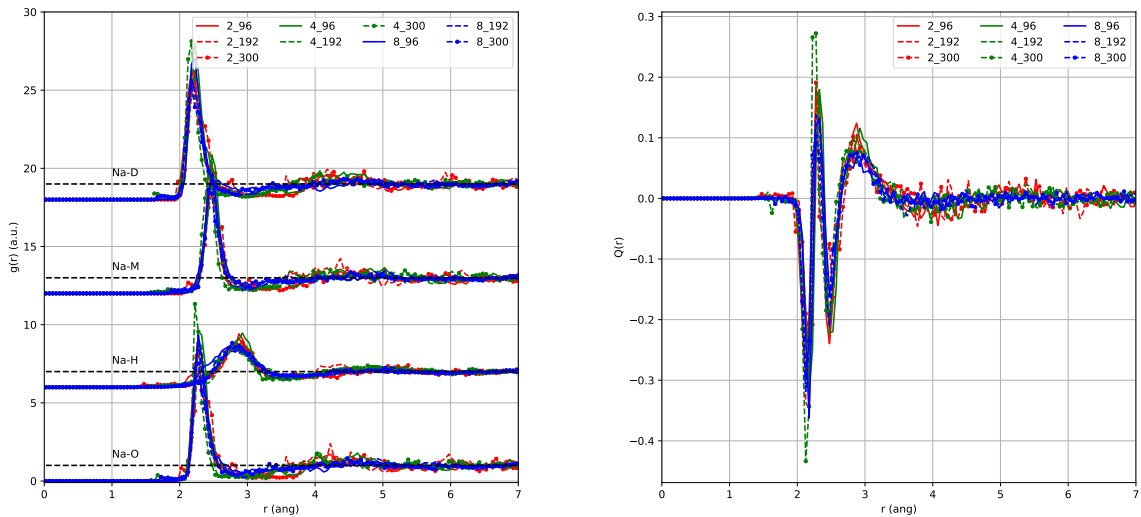

	\centering
	\includegraphics[scale=0.4]{./na_box_rdfs.pdf}
	\includegraphics[scale=0.4]{./na_box_charges.pdf}\\
	\caption{\textbf{(left)} The various RDFs for sodium to SWM4-NDP water molecule constituents. Legend labels correspond to `number of ion pairs'\_`number of waters' in the system. Each set of RDFs is shifted for ease-of-viewing. These RDFs are used to generate the \textbf{(right)} induced charge around the sodium ion in each of these simulations.}\label{supp-fig:na_rdfs_qs}
\end{figure}

\begin{figure}
    \centering
    \includegraphics[scale=0.4]{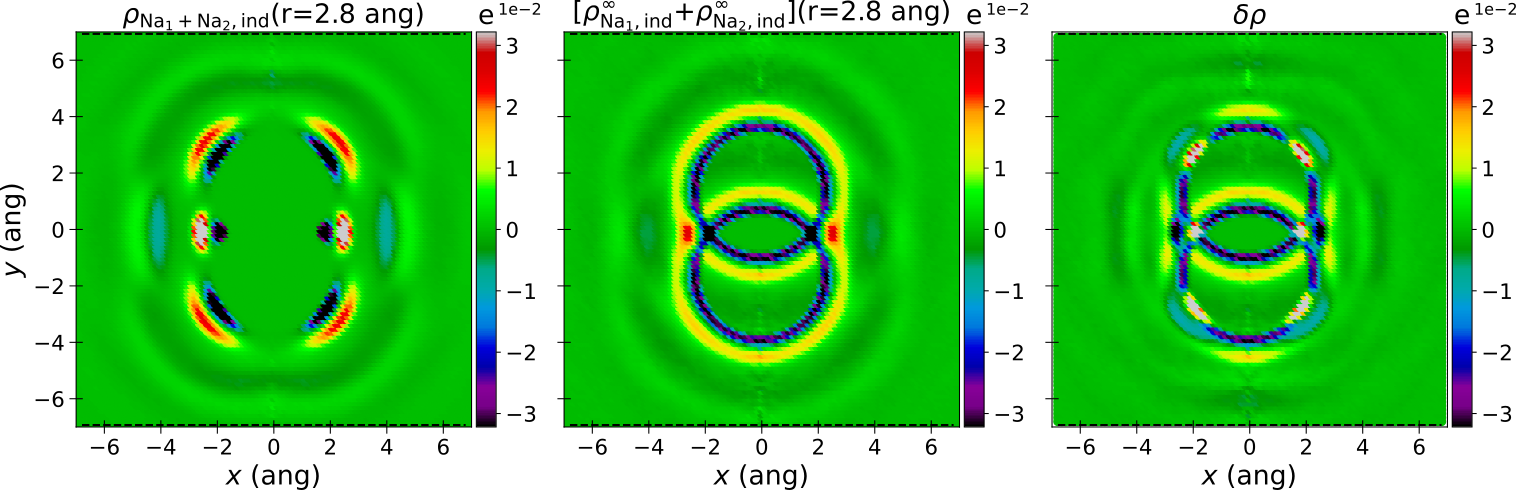}\\
    \vspace{0.2cm}
    \includegraphics[scale=0.4]{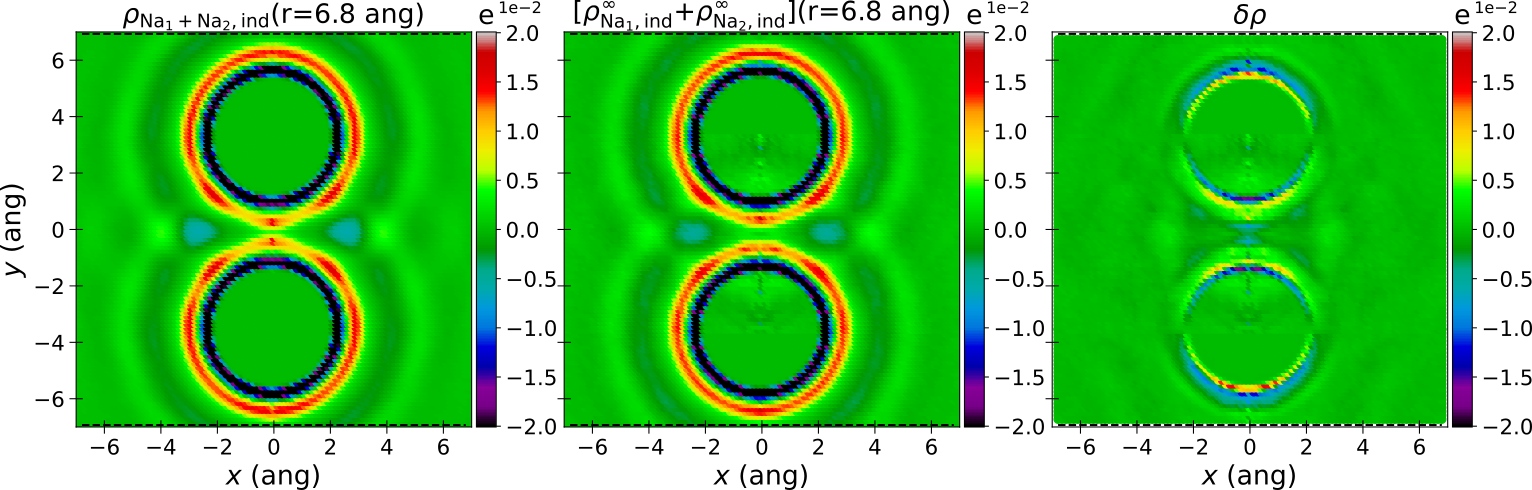}
    
    \caption{\textbf{(left)} The two-dimensional ion-pair simulation charge distributions given by Eq. 8 alongside the \textbf{(middle)} reconstructed charge distribution from the infinitely-dilute ion simulations and \textbf{(right)} the deviations between the two for the case when \textbf{(top)} the interionic separation $r=2.8$ \AA\ and \textbf{(bottom)} when $r=6.8$ \AA. Notably, the $\delta \rho$ at $r=2.8$ \AA\ is an order of magnitude larger than at $r=6.8$ \AA, in agreement with the results presented in Fig. 7.}\label{supp-fig:nanasdfs}
\end{figure}
\twocolumngrid

 \begin{figure*}[t!]
     \centering
     \includegraphics[scale=0.75]{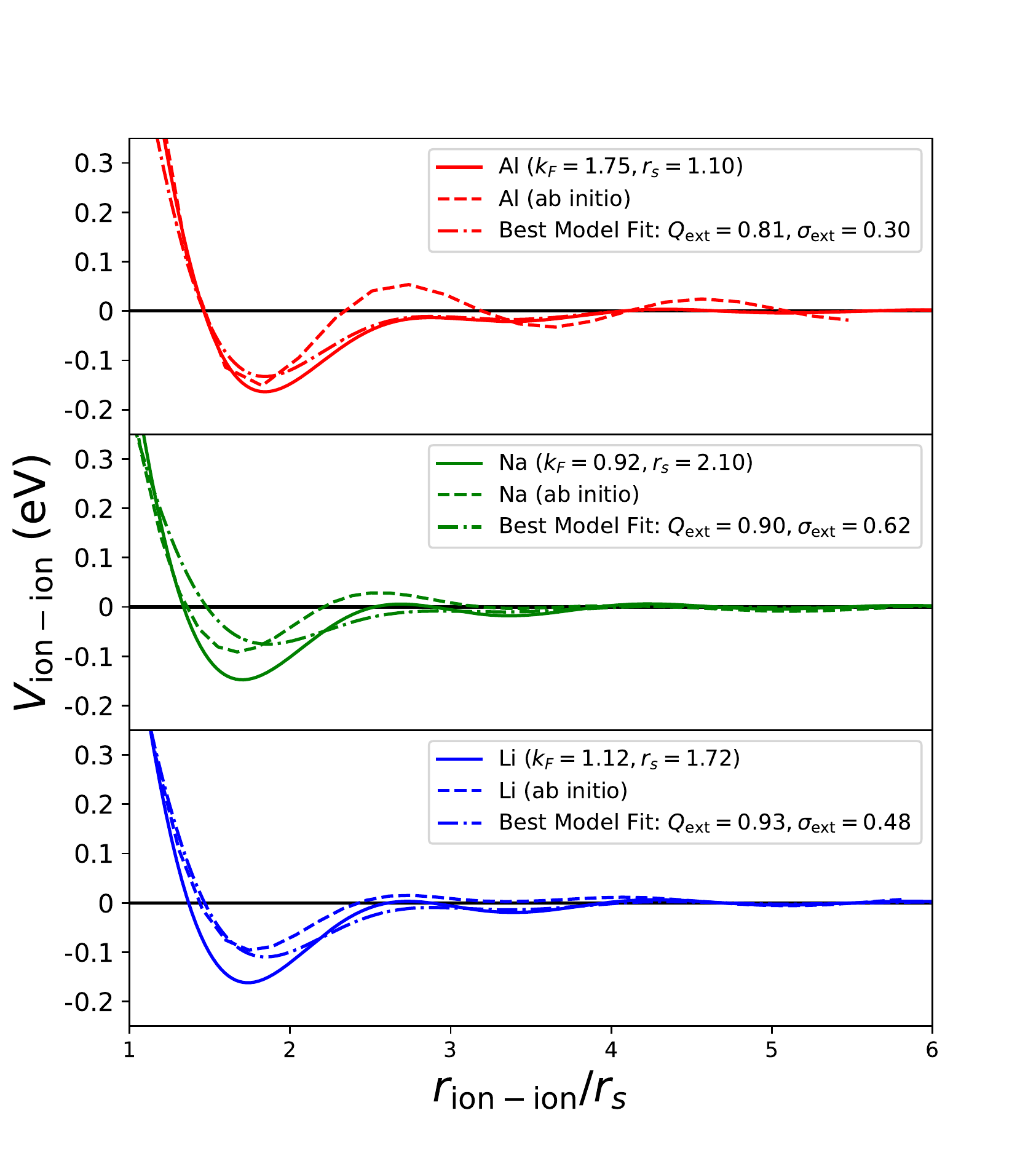}
     \caption{A comparison between our model predictions (solid), the binding curves from the \textit{ab initio} metal simulations (dashed), and model fits to \textit{ab initio} results (dash-dot) for (\textbf{top}) an aluminum lattice ($k_F = 1.75$ \AA$^{-1}$, $r_s = 1.10$ \AA), (\textbf{middle}) a sodium lattice ($k_F = 0.92$ \AA$^{-1}$, $r_s = 2.10$ \AA), and (\textbf{bottom}) a lithium lattice ($k_F = 1.12$ \AA$^{-1}$, $r_s = 1.72$ \AA). We see that the fits correspond to almost fully-ionized protons in the metal, with charge extents on the order of the hydrogen atom.}
     \label{supp-fig:model_fits}
\end{figure*}